\documentclass[twocolumn,twocolappendix]{aastex631}
\usepackage{amsmath}

\shortauthors{Behmard et al.}
\graphicspath{{./}{figures/}}

\begin{document}

\title{Planet Engulfment Detections are Rare According to Observations and Stellar Modeling}

\author[0000-0003-0012-9093]{Aida Behmard}
\affiliation{Division of Geological and Planetary Sciences, California Institute of Technology, Pasadena, CA 91125, USA}

\author[0000-0002-8958-0683]{Fei Dai}
\affiliation{Department of Astronomy, California Institute of Technology, Pasadena, CA 91125, USA}

\author[0000-0002-9873-1471]{John M. Brewer}
\affiliation{Department of Physics and Astronomy, San Francisco State University, 1600 Holloway Avenue, San Francisco, CA 94132, USA}

\author[0000-0002-2580-3614]{Travis A. Berger}
\affiliation{Exoplanets and Stellar Astrophysics Laboratory, Code 667, NASA Goddard Space Flight Center, Greenbelt, MD, 20771, USA}

\author[0000-0001-8638-0320]{Andrew W. Howard}
\affiliation{Department of Astronomy, California Institute of Technology, Pasadena, CA 91125, USA}



\begin{abstract}
Dynamical evolution within planetary systems can cause planets to be engulfed by their host stars. Following engulfment, the stellar photosphere abundance pattern will reflect accretion of rocky material from planets. Multi-star systems are excellent environments to search for such abundance trends because stellar companions form from the same natal gas cloud and are thus expected to share primordial chemical compositions to within 0.03$-$0.05 dex. Abundance measurements have occasionally yielded rocky enhancements, but few observations targeted known planetary systems. To address this gap, we carried out a Keck-HIRES survey of 36 multi-star systems where at least one star is a known planet host. We found that only HAT-P-4 exhibits an abundance pattern suggestive of engulfment, but is more likely primordial based on its large projected separation (30,000 $\pm$ 140 AU) that exceeds typical turbulence scales in molecular clouds. To understand the lack of engulfment detections among our systems, we quantified the strength and duration of refractory enrichments in stellar photospheres using \texttt{MESA} stellar models. We found that observable signatures from 10 $M_{\oplus}$ engulfment events last for $\sim$90 Myr in 1 $M_{\odot}$ stars. Signatures are largest and longest lived for 1.1$-$1.2 $M_{\odot}$ stars, but are no longer observable $\sim$2 Gyr post-engulfment. This indicates that engulfment will rarely be detected in systems that are several Gyr old. 



\end{abstract}



\section{Introduction} \label{sec:intro}
Gravitationally bound stars form from the approximately homogeneous material of their shared natal gas cloud; it follows that differences in their elemental abundances are expected to fall within the small range of chemical dispersion observed in stellar clusters and associations (e.g., \citealt{de_silva2007,de_silva2009,bland_hawthorn2010}). However, several studies have found abundance differences $>$0.05 dex\footnote{In this work, we adopt the standard ``bracket" chemical abundance notation [X/H] = $A$(X) - $A$(X)$_{\odot}$, where $A$(X) = log($n_{\textrm{X}}$/$n_{H}$) + 12 and $n_{\textrm{X}}$ is the number density of species X in the star's photosphere.} between stars in binary systems  \citep{ramirez2011,mack2014,tucci_maia2014,teske2015,ramirez2015,biazzo2015,saffe2016,teske2016,adibekyan2016,saffe2017,tucci_maia2019,ramirez2019,nagar2020,galarza2021,jofre2021}, with extreme cases exhibiting differences up to $\sim$0.2 dex \citep{oh2018}. 

There are various proposed mechanisms for these abundance differences related to planet formation. For example, observed refractory element depletion can be attributed to missing solid material locked up in rocky planets. \citet{melendez2009} put forward this scenario to explain the Sun's observed depletion pattern, but noted that it only makes sense if the combined mass of the Solar System terrestrial planets is removed from just the solar convective zone. It is possible that dust-depleted gas was accreted onto the Sun 10$-$25 Myr after Solar System formation, once the solar convective zone began shrinking to its current mass fraction ($\sim$2\%, \citealt{hughes2007}). However, only 1\% of stars with ages $\geq$13 Myr show signs of accretion \citep{white2005,currie2007}, indicating that late-stage accretion after the protoplanetary disk has dissipated (typical lifetimes 1$-$3 Myr, \citealt{li2016}) is rare. Thus, we do not expect that sequestration of refractory material in planets will produce strong depletion signals. Alternatively, \citet{booth2020} suggested that depletion trends may emerge from gaps in protoplanetary disks created by forming giant planets. These gaps could create pressure traps that prevent accretion of refractory material onto the host star, perhaps from Late Heavy Bombardment-like events. 

Abundance differences can also be produced from refractory enrichment. A particularly promising scenario for producing strong enrichment signals is planet engulfment, which could deposit large amounts of rocky planetary material within the convective regions of engulfing stars. Spectral analysis of polluted white dwarfs provide strong evidence for planet engulfment (e.g., \citealt{zuckerman2010,koester2014,farihi2016}), with some white dwarfs exhibiting surface abundance patterns that closely match bulk Earth composition material (e.g., \citealt{zuckerman2007,klein2010}). There is also evidence for planet engulfment in solar-like stars. For example, \citet{oh2018} recently reported a strong ($\sim$0.2 dex) potential signature of planet engulfment in the HD 240429-30 (Kronos-Krios) system. We investigate abundance differences between stellar companions through the lens of planet engulfment here. 


There are ten binary systems reported in the literature with one star significantly enhanced in refractories ($>$0.05 dex) compared to its stellar companion. Among these ten systems, seven host known planets \citep{ramirez2011,mack2014,tucci_maia2014,teske2015,ramirez2015,biazzo2015,teske2016,saffe2017,tucci_maia2019,jofre2021}. Depending on the study, four to seven of these planet host systems have refractory differences that trend with condensation temperature $T_c$ (Table \ref{tab:table1}). We expect a $T_{c}$-dependent differential abundance pattern following planet engulfment; in the absence of engulfment, elements with higher $T_{c}$ are more likely to be condensed throughout the disk and become locked in solid planetary material. Conversely, elements with lower $T_{c}$ are more likely to reside in the gas phase and become depleted through accretion onto the host star. Thus, rocky planetary compositions are dictated by the radial temperature gradient in the disk, with higher abundances of refractory species in order of $T_c$. Additionally, a $T_{c}$-dependent differential abundance pattern is not expected to result from stellar processes alone.

\begin{figure} 
\centering
    \vspace*{0.02in}
    \includegraphics[width=0.47\textwidth]{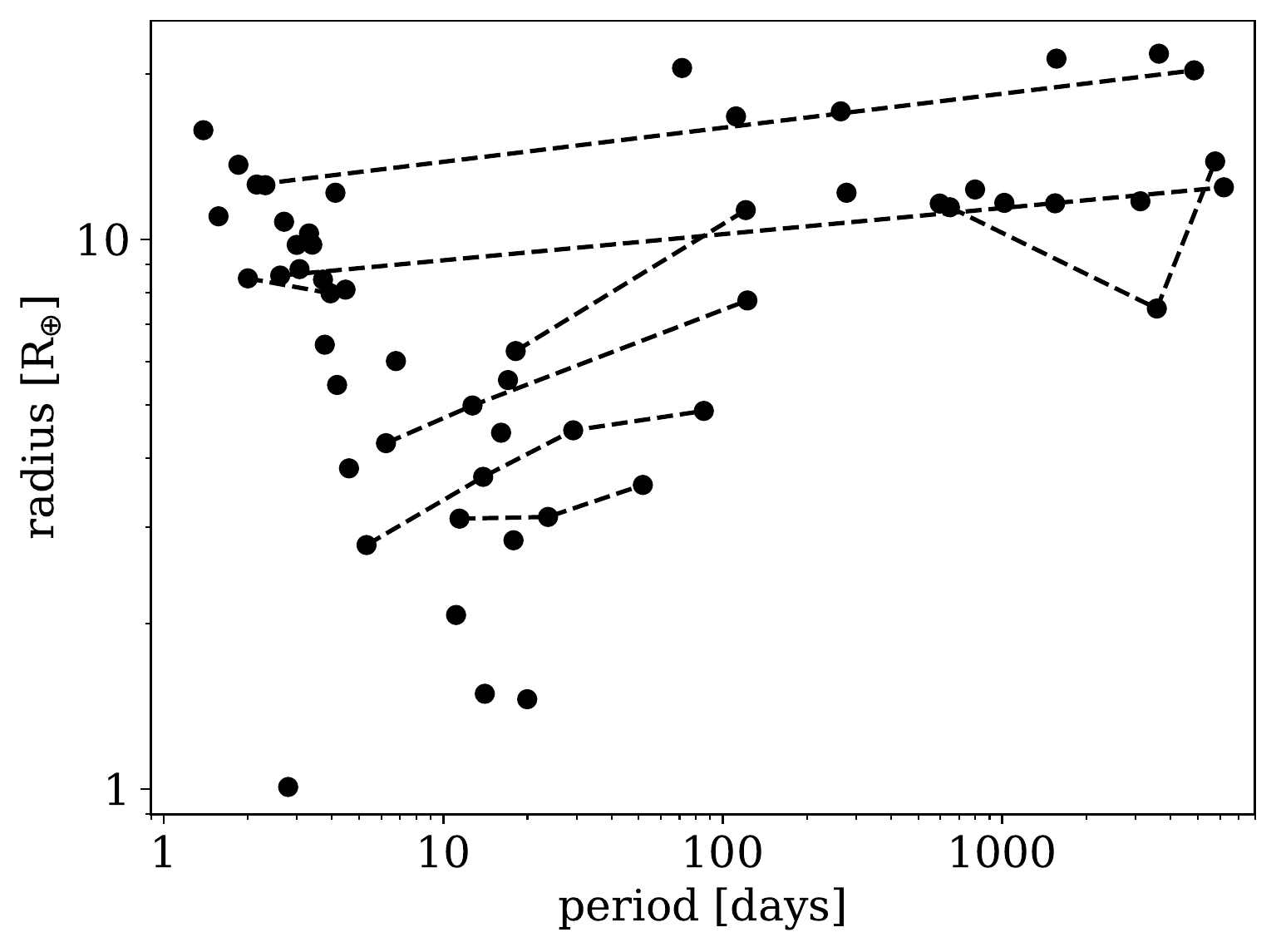}
    \caption{The radii vs. orbital period distribution for planets in our sample. Hot/warm Jupiters are defined as planets with $R$ $>$ 8 $R_{\oplus}$ and $P$ $<$ 100 days, hot/warm sub-Saturns with 4 $R_{\oplus}$ $<$ $R$ $<$ 8 $R_{\oplus}$ and $P$ $<$ 100 days, cold Jupiters with $R$ $>$ 8 $R_{\oplus}$ and $P$ $>$ 100 days, and super-Earths/sub-Neptunes with $R$ $<$ 4 $R_{\oplus}$. Planets that share the same host star are connected by dashed lines.}
\label{fig:figure1}
\end{figure}

There have been a few differential abundance studies for larger samples. For example, \citet{hawkins2020} reported abundances for 25 comoving, wide binaries and found that while 80\% (20 pairs) are homogeneous in [Fe/H] at levels below 0.02 dex, the five remaining systems exhibit $\Delta$[Fe/H] $\sim$ 0.10 dex. If we assume that these refractory enhancements indicate planet engulfment, they imply an engulfment rate of 20\%. However, the authors did not recover a strong $T_{c}$ trend for any of the $\Delta$[Fe/H] $\sim$ 0.10 dex systems, suggesting that the abundance differences may stem from other processes. The absence of a strong $T_{c}$ trend could also be attributed to a lack of low $T_{c}$ element measurements in the \citet{hawkins2020} sample, which makes the $T_{c}$ trend difficult to discern, or abundance measurement error. More recently, \citet{spina2021} analyzed differential abundances among 107 binary systems. While they did not assess $T_{c}$ trends, they found that $\sim$20$-$35\% of their sample exhibits large refractory-to-volatile abundance ratios that may be indicative of engulfment. While these results are intriguing, they highlight the need for further high-precision abundance studies that consider $T_c$ to constrain the true rate of planet engulfment.

\begin{deluxetable*}{lrcrcrrl}
\setlength{\tabcolsep}{0.25em}
\tablewidth{0.9\textwidth}
\tabletypesize{\footnotesize}
\tablewidth{0pt}
\tablecaption{Planet Host Binaries with Previously Measured High-Precision Abundances \label{tab:table1}}
\tablecolumns{8}
\tablehead{
\colhead{System} &
\colhead{$|\Delta T_{\textrm{eff}}|$} &
\colhead{$|\Delta$log$g|$} &
\colhead{sep} &
\colhead{$|\Delta$[Fe/H]$|$} &
\colhead{Instrument} &
\colhead{$T_c$ trend} &
\colhead{Source} \\[-0.2cm]
\colhead{} &
\colhead{K} &
\colhead{dex} &
\colhead{AU} &
\colhead{dex} &
\colhead{} &
\colhead{} &
\colhead{}
}
\startdata
HAT-P-1 & 17  & 0.07  & 1550  & 0.009 $\pm$ 0.009 & Keck-HIRES$^{\textrm{a}}$ & no & \citet{liu2014} \\
HD 20781-82  & 465  & 0.10    & 9000 & 0.060 $\pm$ 0.010 & Magellan-MIKE$^{\textrm{b}}$ & yes & \citet{mack2014} \\
XO-2    & 60   & 0.02   & 4500  & 0.054 $\pm$ 0.005 & Subaru-HDS$^{\textrm{c}}$;  & maybe/yes & \citet{teske2015}; \\
& & & & & Keck-HIRES$^{\textrm{a}}$ & & \citet{ramirez2015}; \\
  &  &  &  &  & HARPS-N$^{\textrm{d}}$ &  & \citet{biazzo2015} \\
WASP-94  & 82  & 0.09   & 2700   & 0.015 $\pm$ 0.004 & Magellan-MIKE$^{\textrm{b}}$ & yes &\citet{teske2016} \\
HAT-P-4 & 10  & 0.05  & 28446  & 0.105 $\pm$ 0.006 & Gemini-GRACES$^{\textrm{e}}$ & yes & \citet{saffe2017} \\
HD 80606-07  & 52  & 0.04   & 1200  & 0.000 $\pm$ 0.040 & Keck-HIRES$^{\textrm{a}}$ & no & \citet{saffe2015,mack2016}; \\ & & & & & & & \citet{liu2018} \\
16 Cygni  & 79  & 0.05    & 860   & 0.047 $\pm$ 0.005 & McDonald-RGT$^{\textrm{f}}$ & yes/no & \citet{ramirez2011};
\\ & & & & & CFHT-ESPaDOnS$^{\textrm{g}}$ & & \citet{tucci_maia2014}; \\
 & & & & & Subaru-HDS$^{\textrm{c}}$ & &  \citet{tucci_maia2019} \\ & & & & & & & \citet{ryabchikova2022} \\
HD 133131  & 5  & 0.0   & 360   & 0.032 $\pm$ 0.015 & Magellan-MIKE$^{\textrm{b}}$, VLT-UVES$^{\textrm{h}}$ & maybe &\citet{teske2016b,liu2021} \\
HD 106515  & 250  & 0.09    & 860   & 0.008 $\pm$ 0.01 & Keck-HIRES$^{\textrm{a}}$, VLT-UVES$^{\textrm{h}}$ & no & \citet{saffe2019,liu2021} \\
WASP-160  & 8060  & 0.05    & 860   & 0.012 $\pm$ 0.017 & Gemini-GRACES$^{\textrm{e}}$ & yes & \citet{jofre2021}
\enddata
\tablecomments{(a) Keck High Resolution Echelle Spectrograph \citep{vogt1994}, (b) Magellan Inamori Kyocera Echelle high-resolution spectrograph \citep{bernstein2003}, (c)
High Dispersion Spectrograph \citep{noguchi2002}, (d) Telescopio Nazionale Galileo HARPS-N Spectrograph \citep{cosentino2012}, (e) Gemini Remote Access to CFHT ESPaDOnS Spectrograph \citep{chene2014}, (f) McDonald Observatory R.G. Tull spectrograph \citep{tull1972}, (g) Canada-France-Hawaii Telescope ESPaDOnS spectrograph \citep{petit2003}, (h) ESO Very Large Telescope UV-visual echelle spectrograph \citep{dekker2000}}
\end{deluxetable*}

Understanding the conditions and prevalence of planet engulfment is vital for mapping the fate of refractory material within planetary systems. There are multiple lines of evidence that solid planetary material is predominantly refractory. For example, white dwarf pollution patterns from planet debris exhibit rocky compositions \citep{xu2019,putirka2021}, and the bulk densities of several super-Earth exoplanets, e.g., the TRAPPIST-1 planets and Kepler-93b \citep{dressing2015}, are indicative of Earth-like rock-iron ratios. Thus, the building blocks of planets are sourced from the dusty component of protoplanetary disks. However, it is not clear how much disk dust becomes locked in planets or sequestered in debris disks (e.g., \citealt{booth2020}), is engulfed by the host star following a combination of radial drift and dynamical interactions, or is blown out of the system. In other words, we have not quantified the efficiency of planet formation. Refractory enhancements in planet host stars due to engulfment can be used to back out mass measurements of polluting refractory material, which will shed light on how much mass went into planets or was trapped in the outer disk, and how that mass was redistributed in the system after the disk dissipated. 

The prevalence of planet engulfment also has implications for stellar chemical evolution. Stars are born together in clusters, but disperse over time. Galactic archaeology attempts to link stars back to their siblings through chemical tagging that can trace the chemical and kinematic evolution of the Milky Way. However, chemical tagging relies on the assumption that such stellar siblings are coeval and share the same elemental abundance patterns to within 0.03$-$0.05 dex (e.g., \citealt{de_silva2007,bovy2016,ness2018}). This assumption may not be true if planet engulfment is a common phenomenon. Indeed, it has been suggested that observations of significant chemical dispersion observed within stellar clusters and associations, such as inhomogeneities in neutron capture elements within the open cluster M67 \citep{liu2016a}, and abundance differences at the 0.02 dex level for 19 elements in the Hyades open cluster \citep{liu2016b}, are due to planet engulfment \citep{oh2017,ness2018}. 

In addition, there are no high-precision abundance surveys that specifically targeted planet hosts. Assessing engulfment signatures in systems with existing planets is important for understanding the dynamical conditions that may give rise to planet engulfment, such as planet-planet scattering in multi-planet systems \citep{rasio1996,weidenschilling1996}. 
To fill this gap, we carried out a survey with the Keck High Resolution Echelle Spectrometer (HIRES) of 36 confirmed planet host systems with stellar companions to investigate the role of engulfment in planetary system evolution, and shed light on which dynamical pathways may dominate. For more details on the sample, see Section \ref{sec:sample}. The abundance analysis and engulfment model used to derive mass measurements of engulfed material are presented in Sections \ref{sec:abundances} and \ref{sec:model}, respectively. Our \texttt{MESA} analysis is outlined in Section \ref{sec:timescales}. The results of our survey are presented in Section \ref{sec:potential_detections}, and are compared to previously published results in Section \ref{sec:previous_systems}. Implications for planet engulfment and chemical homogeneity in multi-star systems are discussed in Section \ref{sec:discussion}. Finally, we summarize our findings in Section \ref{sec:summary}.



\begin{deluxetable*}{lrrrrlrrrrr}
\setlength{\tabcolsep}{1em}
\tablewidth{0.9\textwidth}
\tabletypesize{\footnotesize}
\tablewidth{0pt}
\tablecaption{Observational Properties of Engulfment Sample Stars \label{tab:table2}}
\tablecolumns{11}
\tablehead{
\colhead{Name} &
\colhead{RA} &
\colhead{Dec} &
\colhead{$T_{\textrm{eff}}$} &
\colhead{log$g$} &
\colhead{$M_{*}$} &
\colhead{RV} &
\colhead{$\pi$} &
\colhead{$\mu_{\alpha}$} &
\colhead{$\mu_{\delta}$} &
\colhead{G} \\
\colhead{} &
\colhead{deg:mm:ss} &
\colhead{deg:mm:ss} &
\colhead{K} &
\colhead{dex} &
\colhead{$M_{\odot}$} &
\colhead{km s$^{-1}$} &
\colhead{mas} &
\colhead{mas yr$^{-1}$} &
\colhead{mas yr$^{-1}$} &
\colhead{mag} 
}
\startdata
       HAT-P-4 A* &  15:19:57.89 &   36:13:46.35 &      5903 &        4.14 &     1.31 &  $-$1.67 &             3.11 &       $-$21.51 &        $-$24.25 &  11.1 \\
      HAT-P-4 B &  15:19:59.98 &   36:12:18.13 &      5919 &        4.17 &    1.10$^{\dagger}$ &   $-$1.94 &             3.08 &       $-$21.42 &        $-$24.18 &  11.4 \\
    HD 132563 A &  14:58:21.43 &    44:02:34.25 &      6158 &        4.18 &     1.16 &    -- &             9.41 &       $-$62.79 &        $-$67.68 &   8.9 \\
   HD 132563 B* &  14:58:21.05 &    44:02:34.74 &      6032 &        4.32 &    1.09 &   $-$5.98 &             9.47 &       $-$57.48 &        $-$70.15 &   9.3 \\
   HD 133131 A* &    15:03:36.00 &  $-$27:50:29.81 &      5827 &        4.50   & 0.86 &   $-$15.34 &            19.41 &       159.01 &       $-$139.13 &   8.3 \\
   HD 133131 B* &   15:03:35.63 &  $-$27:50:35.36 &      5815 &        4.48   & 0.86 &   $-$16.63 &            19.43 &       156.23 &       $-$133.77 &   8.3 \\
   $\omega$ Ser A* &  15:50:17.58 &    02:11:46.67 &      4900 &        2.87   & 1.97 &    $-$3.65 &            13.10 &        30.26 &        $-$47.59 &   4.9 \\
    $\omega$ Ser B &  15:50:13.27 &    02:12:24.42 &      5252 &        4.54   & 0.88$^{\dagger}$ &    $-$3.24 &            12.90 &        30.55 &        $-$48.53 &  10.1 \\
    HD 178911 A &    19:09:04.45 &    34:36:04.56 &      5849 &        4.20   & 1.39 &   $-$38.09 &            20.23 &        76.62 &        207.13 &   6.6 \\
   HD 178911 B* &    19:09:03.17 &    34:36:02.61 &      5563 &        4.39   & 1.03 &      -- &            24.41 &        57.18 &        195.81 &   7.9 \\
       16 Cyg A &  19:41:48.71 &   50:31:27.68 &      5781 &        4.28 &  1.02 &    $-$27.21 &            47.32 &      $-$148.03 &       $-$159.03 &   5.8 \\
      16 Cyg B* &  19:41:51.75 &    50:31:00.49 &      5746 &        4.37 &   0.98 &  $-$27.73 &            47.33 &      $-$134.48 &       $-$162.70 &   6.1 \\
   HD 202772 A* &   21:18:47.90 &  $-$26:36:58.98 &      6255 &        3.91 &  1.48 &   $-$17.71 &             6.14 &        23.25 &        $-$57.67 &   8.2 \\
    HD 202772 B &  21:18:47.81 &  $-$26:36:58.44 &      6103 &        4.14 &  1.26 &      -- &             6.33 &        28.91 &        $-$56.51 &  10.0 \\
    HAT-P-1 A &  22 57 45.96 &   38:40:26.53 &      6069 &        4.12 &  1.23$^{\dagger}$ &    $-$3.02 &             6.24 &        32.08 &        $-$42.08 &   9.6 \\
     HAT-P-1 B* &  22:57:46.89 &   38:40:29.69 &      5966 &        4.32 &  1.13 &    $-$2.98 &             6.24 &        32.42 &        $-$41.95 &  10.2 \\
   Kepler-25 A* &   19:06:33.21 &   39:29:16.46 &      6214 &        4.12 &  1.14 &    $-$7.59 &             4.15 &        $-$0.30 &          6.11 &  10.6 \\
    Kepler-25 B &   19:06:32.52 &    39:29:19.10 &      4825 &        4.47 &   0.80 &     -- &             4.11 &         0.32 &          6.18 &  13.2 \\
     WASP-94 A* &   20:55:07.98 &    $-$34:08:08.73 &      6042 &        4.16 &  1.36 &    $-$8.30 &             4.75 &        26.50 &        $-$44.97 &  10.0 \\
      WASP-94 B &   20:55:09.19 &    $-$34:08:08.63 &      5987 &        4.23 & 1.24 &     $-$8.45 &             4.72 &        26.19 &        $-$44.70 &  10.4 \\
      HD 20781* &    03:20:03.37 &   $-$28:47:02.86 &      5232 &        4.45 & 0.86 &     40.31 &            27.81 &       348.87 &        $-$66.61 &   7.2 \\
      HD 20782* &     03:20:04.00 &  $-$28 51 15.71 &      5760 &        4.36 & 0.93 &     39.89 &            27.88 &       349.05 &        $-$65.31 &   8.2 \\
    HD 40979 A* &    06:04:30.08 &   44:15:35.15 &      6137 &        4.36 & 1.23 &     32.47 &            29.43 &        95.07 &       $-$152.65 &   6.6 \\
     HD 40979 B &    06:04:13.16 &   44:16:38.63 &      4896 &        4.54 & 0.85 &     33.02 &            29.46 &        94.28 &       $-$153.19 &   8.8 \\
      KELT-2 A* &   06:10:39.37 &   30:57:25.68 &      6142 &        3.96 & 1.48 &    $-$47.22 &             7.43 &        16.73 &         $-$2.15 &   8.6 \\
       KELT-2 B &   06:10:39.28 &   30:57:27.79 &      4847 &        4.41 &  0.80$^{\dagger}$ &      -- &             7.29 &        17.86 &         $-$3.59 &  12.0 \\
    WASP-173 A* &  23:36:40.49 &   $-$34:36:40.70 &      5796 &        4.49 &  1.10 &      -- &             4.24 &        87.91 &         $-$8.71 &  11.4 \\
     WASP-173 B &  23:36:40.96 &  $-$34:36:42.82 &      5441 &        4.43 &   0.95$^{\dagger}$ &     -- &             4.27 &        87.41 &         $-$8.95 &  12.0 \\
    WASP-180 A* &   08:13:34.14 &   $-$01:58:58.04 &      6316 &        4.41 &   1.22 &   27.73 &             3.98 &       $-$13.89 &         $-$2.82 &  10.9 \\
     WASP-180 B &   08:13:34.35 &     $-$01:59:01.70 &      5808 &        4.53 &  1.07 &    27.99 &             3.84 &       $-$13.23 &         $-$2.79 &  11.8 \\
  Kepler-515 A* &  19:21:58.64 &    52:03:18.98 &      5197 &        4.52 &  0.80 &    $-$8.05 &             3.05 &       $-$23.45 &        $-$71.26 &  13.2 \\
   Kepler-515 B &  19:21:58.42 &    52:03:19.08 &      4798 &        4.52 &  0.71 &      -- &             3.07 &       $-$24.18 &        $-$71.90 &  13.8 \\
  Kepler-477 A &  19:12:16.16 &   42:21:18.66 &      4921 &        4.52 &    0.74 &    -- &             2.12 &       $-$28.65 &        $-$11.31 &  14.1 \\
  Kepler-477 B* &  19:12:16.22 &   42:21:19.68 &      5177 &        4.57 &   0.68 &     -- &             2.15 &       $-$29.12 &        $-$11.96 &  14.5 \\
 Kepler-1063 A* &   19:22:06.43 &    38:08:34.15 &      5568 &        4.38 &  1.02 &      -- &             1.93 &        $-$8.54 &        $-$14.90 &  12.9 \\
  Kepler-1063 B &   19:22:06.37 &    38:08:34.98 &      5783 &        4.35 &  1.12 &      -- &             1.86 &        $-$9.19 &        $-$15.16 &  13.2 \\
      WASP-3 A* &  18:34:31.62 &   35:39:41.14 &      6319 &        4.17 &  1.20 &    $-$4.40 &             4.33 &        $-$5.79 &        $-$21.93 &  10.5 \\
       WASP-3 C &  18:34:30.25 &   35:39:33.63 &      4553 &        4.43 &   0.70 &     -- &             4.28 &        $-$7.54 &        $-$23.29 &  13.6 \\
    WASP-160 A &   05:50:44.74 &   $-$27:37:05.68 &      5155 &        4.46 &  0.94 &    $-$6.03 &             3.46 &        26.87 &        $-$34.75 &  12.5 \\
    WASP-160 B* &    05:50:43.10 &  $-$27:37:23.98 &      5370 &        4.40 & 0.92 &     $-$6.08 &             3.45 &        27.03 &        $-$34.80 &  12.9 \\
      HD 80606* &   09:22:37.67 &    50:36:13.60 &      5523 &        4.32 & 1.05 &      4.16 &            15.14 &        56.02 &         10.33 &   8.8 \\
       HD 80607 &   09:22:39.83 &   50:36:14.11 &      5475 &        4.32 &  1.03 &     3.70 &            15.15 &        52.66 &          9.94 &   9.0 \\
         XO-2N* &    07:48:06.42 &   50:13:30.45 &      5272 &        4.31 & 0.98 &     47.68 &             6.66 &       $-$29.55 &       $-$154.23 &  11.0 \\
         XO-2S* &    07:48:07.43 &    50:13:00.79 &      5273 &        4.32 & 1.00 &     46.85 &             6.67 &       $-$29.31 &       $-$154.23 &  10.9 \\
       HD 99491 &  11:26:44.55 &     03:00:50.05 &      5431 &        4.38 &  1.02 &     3.96 &            55.01 &      $-$725.96 &        180.98 &   6.3 \\
      HD 99492* &   11:26:45.50 &     03:00:25.77 &      4898 &        4.45 &   0.86 &    3.51 &            55.06 &      $-$728.13 &        188.55 &   7.3 \\
   HD 106515 A* &    12:15:06.30 &   $-$07:15:27.17 &      5371 &        4.41 &  0.90 &     20.82 &            29.31 &      $-$251.47 &        $-$51.33 &   7.7 \\
    HD 106515 B &   12:15:05.84 &   $-$07:15:27.67 &      5220 &        4.41 &  0.88 &    19.99 &            29.39 &      $-$244.60 &        $-$67.74 &   8.0 \\
      WASP-64 A &    06:44:29.50 &  $-$32:51:29.49 &      5770 &        4.24 &  1.04 &    35.48 &             2.76 &       $-$19.41 &         $-$1.90 &  11.3 \\
     WASP-64 B* &   06:44:27.58 &   $-$32:51:30.20 &      5691 &        4.44 &  1.37$^{\dagger}$ &    35.06 &             2.77 &       $-$19.27 &         $-$1.07 &  12.5 \\
    WASP-127 A* &   10:42:14.10 &    $-$03:50:05.99 &      5824 &        4.21 &  0.93 &    $-$8.25 &             6.22 &        19.13 &         17.06 &  10.1 \\
     WASP-127 B &  10:42:11.44 &   $-$03:50:12.78 &      5566 &        4.50 & 0.95$^{\dagger}$ &     $-$8.19 &             6.21 &        18.77 &         16.49 &  11.1 
   \enddata
   \tablecomments{This is a subset of a table that lists the equatorial coordinates, $T_{\textrm{eff}}$, log$g$, $M_{*}$, \emph{Gaia} EDR3-sourced radial velocities, parallaxes, proper motions, and $G$-magnitudes for stars in the engulfment sample. $T_{\textrm{eff}}$ and log$g$ were calculated by applying \texttt{SME} to the Keck-HIRES spectra. $M_{*}$ were generated via \texttt{SpecMatch-Syn} \citep{petigura15}, except for targets marked with $^{\dagger}$, which were obtained from \citet{mugrauer2019} or the \emph{NASA Exoplanet Archive}$^{\ref{footnote 2}}$. The brighter component of each binary pair is denoted as $`$A', and the fainter component as $`$B'. The planet hosts are marked with *.
   \vspace{1mm}
   \newline (This table is available in its entirety in machine-readable form.)}
\end{deluxetable*}

\section{Planet Engulfment Sample} \label{sec:sample}
Our planet engulfment sample consists of multi-star systems where at least one star is a confirmed planet host. The sample is largely sourced from the \citet{mugrauer2019} catalog of 207 confirmed planet hosts with stellar companions at separations of $<$9100 AU, compiled from the second data release of the \emph{Gaia} mission (\emph{Gaia} DR2, \citealt{gaia_dr2}). The companions were identified through a set of astrometric conditions that when met constitute strong evidence that a pair of stars are gravitationally bound. For more details on the companion selection criteria, see \citet{mugrauer2019}.

We applied a projected separation cut of $>$1.5$''$ to ensure that the two stars would be cleanly resolved by Keck-HIRES, as well as an effective temperature cut of $T_\textrm{eff}$ = 4700--6500 K. The latter cut was applied because the spectral synthesis code used for our abundance analysis (Spectroscopy Made Easy, \texttt{SME}) does not produce reliable abundances outside of this temperature range (\citealt{valenti1996,brewer2016}). For the companions, we used their $T_\textrm{eff}$ values reported in \citet{mugrauer2019}. These were determined from absolute G-bands magnitudes and the \citet{baraffe2015} (sub)stellar evolution models assuming an age of 5 Gyr, which is the average age of systems in the \citet{mugrauer2019} sample. For the planet hosts, we used the most recently reported $T_\textrm{eff}$ from the \emph{NASA Exoplanet Archive}\footnote{\url{https://exoplanetarchive.ipac.caltech.edu/}\label{footnote 2}}. We foreshadow here that \texttt{SME} provides more accurate $T_\textrm{eff}$ measurements, so this cut was redone after collecting spectra for our targets and running them through \texttt{SME}. This eliminated a further seven systems, which is described in more detail below. However at this point, we were left with 35 systems. We augmented this sample by searching for stellar companions to planet hosts that met these criteria in the \emph{NASA Exoplanet Archive}, which resulted in an additional two systems (HAT-P-4 and WASP-180). Eleven of the 37 planet host binaries qualify as stellar twins ($\Delta$$T_{\textrm{eff}}$ $<$ 200 K, \citealt{andrews2019}), which are well suited to differential abundance analyses given their near-identical evolutionary states. All systems in our sample were verified to host confirmed planets according to the \emph{NASA Exoplanet Archive}$^{\ref{footnote 2}}$. Finally, we removed any systems that display evidence of spectroscopic binary contamination in their spectral cross-correlation; such contamination will lead to inaccurate \texttt{SME} abundance predictions. This was the case for $\psi^1$ Dra, leaving 36 systems.


The final sample of 36 systems contains 28 binaries and eight triples. Though four of the eight triples are hierarchical, we determined that the spectra of individual stars in these systems are not blended with those of nearby companions using the \texttt{ReaMatch} code \citep{kolbl2015}. Each of the triple systems has only one stellar companion that meets the $T_{\textrm{eff}}$ and projected separation criteria. Thus, two stars were always analyzed per system. The equatorial coordinates, $T_{\textrm{eff}}$, log$g$, $M_{*}$, \emph{Gaia} Early Data Release 3 (ER3)-sourced radial velocities, parallaxes, proper motions, and $V$-band magnitudes of stars in the sample are listed in Table \ref{tab:table2}. Some sources are missing radial velocity measurements because they do not meet the Gaia DR2/EDR3 radial velocity criteria of $G$-band magnitudes less than $\sim$13, or were deemed inaccurate due to companion contamination \citep{boubert2019}. Among the 36 systems, ten have existing high-precision abundance measurements (HAT-P-1, HD 20781-82, XO-2, WASP-94, HAT-P-4, HD 80606-07, 16-Cygni, HD 133131, HD 106515, WASP-160; Table \ref{tab:table1}) derived from the \texttt{MOOG} spectral synthesis code \citep{sneden1973,sobeck2011} that can be compared with predictions from \texttt{SME}.

\begin{figure*} 
\centering
    \includegraphics[width=0.98\textwidth]{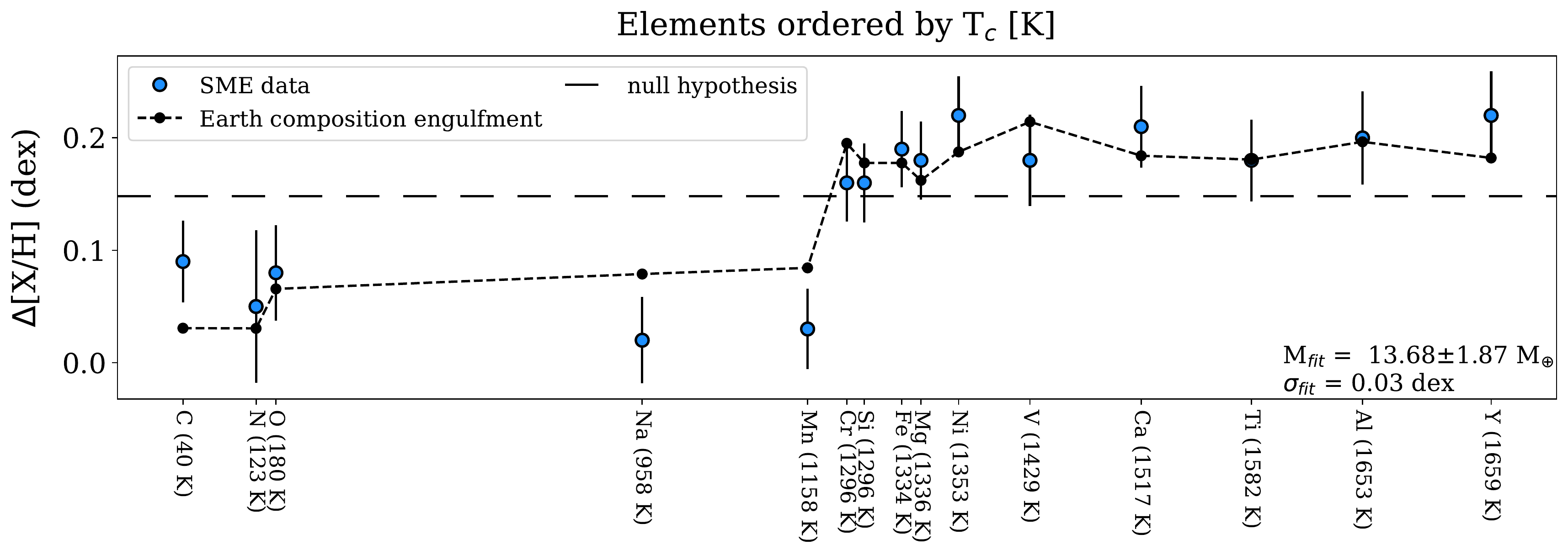}
\caption{Fitted model to the differential abundance measurements between HD 240429 (Krios) and HD 240430 (Kronos) \citep{oh2018}. Blue circles represent the abundance differences from \citet{oh2018}, and black dots are our model fit with 13.68 $\pm$ 1.93 $M_\oplus$ of bulk Earth composition engulfed material added to the convective zone of one star. The abundances are ranked by $T_{c}$ of elements for solar-composition gas from \citet{lodders2003}. The amount of modeled engulfed material and fitted scatter is provided in the lower right corner of the plot. The null hypothesis (no engulfment, but uniform abundance enrichment across all elements for one star) is shown by the long-dash line.} 
\label{fig:figure2}
\end{figure*}

The engulfment sample systems span a wide range of planetary architectures that include super-Earths/sub-Neptunes, compact multi-planet systems, and giant planets at a range of orbital periods (Table \ref{tab:table3}). Figure \ref{fig:figure1} shows the radii versus rotation periods for all planets in the engulfment sample. For planets lacking reported radius measurements, we derived radii from mass measurements with the following power-law mass-radius relation that assumes Earth-like compositions (Rubenzahl et al. \emph{in prep.}):

\begin{equation}
M = CR^{\gamma} 
\end{equation}

\noindent where the $C$ and $\gamma$ were constrained to values of 0.83 and 3.52 using a sample of 122 confirmed exoplanets with Keck-HIRES spectra and precise radii measurements. For planets massive enough to host gaseous envelopes greater than 1\% by mass, the envelope mass was accounted for by assuming a gas density of 0.417 g cm$^{-3}$ as constrained with the Rubenzahl et al. (\emph{in prep.}) planet sample.

\begin{deluxetable}{lr}[h!]
\tablewidth{0.37\textwidth}
\tabletypesize{\footnotesize}
\tablewidth{0pt}
\tablecaption{Sample Architectures \label{tab:table3}}
\tablecolumns{2}
\tablehead{
\colhead{Architecture} &
\colhead{Number} \\
\colhead{} &
\colhead{}
}
\startdata
Hot/Warm Jupiters & 15  \\
Hot/Warm sub-Saturns & 11  \\ 
Cold Jupiters & 15 \\
Cold sub-Saturns & 2 \\
Super-Earths/Sub-Neptunes & 11 
\enddata
\end{deluxetable}

\section{Stellar Abundance Analysis} \label{sec:abundances}

We obtained spectra for these stars with HIRES at the Keck I 10 m telescope \citep{vogt1994} using procedures from the California Planet Search. \citet{howard2010} provides descriptions of the observing and analysis procedures. We used the C2 decker for targets with $V$-band magnitudes fainter than 10 mag, and the B5 decker for targets with $V$-band magnitudes of 10 mag or brighter. The HIRES spectra are high-resolution (R $\approx$ 50,000) with high signal-to-noise ratios per pixel (SNR $\geq$ 40/pixel, with $\sim$50\% having SNR $>$ 100/pixel). The wavelength range utilized spans 350 \AA\ of the spectrum in specific segments between 5164 \AA\ and 7800 \AA, as described in \citet{brewer2016} for their \texttt{SME} implementation. Our choice of SNR $\approx$ 40--400/pix for the engulfment sample HIRES observations was motivated by the expected \texttt{SME} prediction precisions as a function of SNR; for HIRES spectra with SNR = 40--100/pix, \texttt{SME} achieves precisions of 0.01--0.05 dex in [X/H] for the following elements: C, N, O, Na, Mg, Al, Si, Ca, Ti, V, Cr, Mn, Fe, Ni, Li, and Y (e.g., \citealt{brewer2016,brewer2018}). The refractory species alone (Fe, Ti, Al, etc.) achieve higher precision of 0.01--0.03 dex, which translates to detections at the $\sim$1 $M_{\oplus}$ level according to the \citet{oh2018} model used for their analysis of engulfment in the Kronos-Krios system. This precision is sufficient for detecting signatures of planet engulfment, i.e., refractory enhancements, at levels of $>$0.05 dex (e.g., \citealt{ramirez2019}). For reference, an abundance difference of $\Delta$[X/H] = 0.05 dex corresponds to $\sim$2 $M_{\oplus}$ of engulfed solid material assuming a solar-like convective zone mass of $M_{cz}$ = 0.02 $M_{\odot}$ \citep{saffe2017}.


\begin{figure*}[t]
    \centering
    \begin{minipage}{0.455\textwidth}
        \centering
        \includegraphics[width=0.99\textwidth]{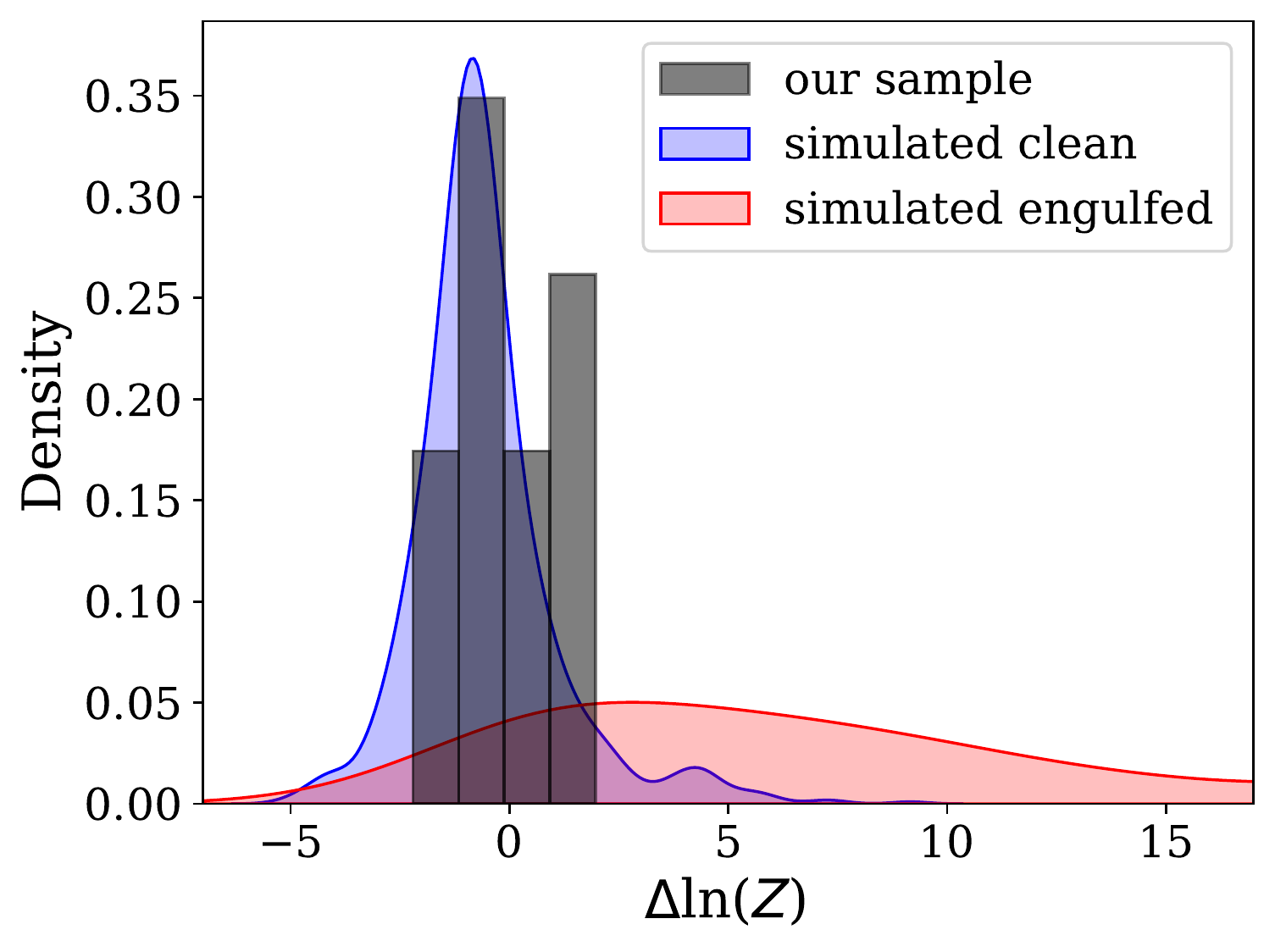} 
    \end{minipage}\hfill
    \begin{minipage}{0.49\textwidth}
        \centering
        \includegraphics[width=0.99\textwidth]{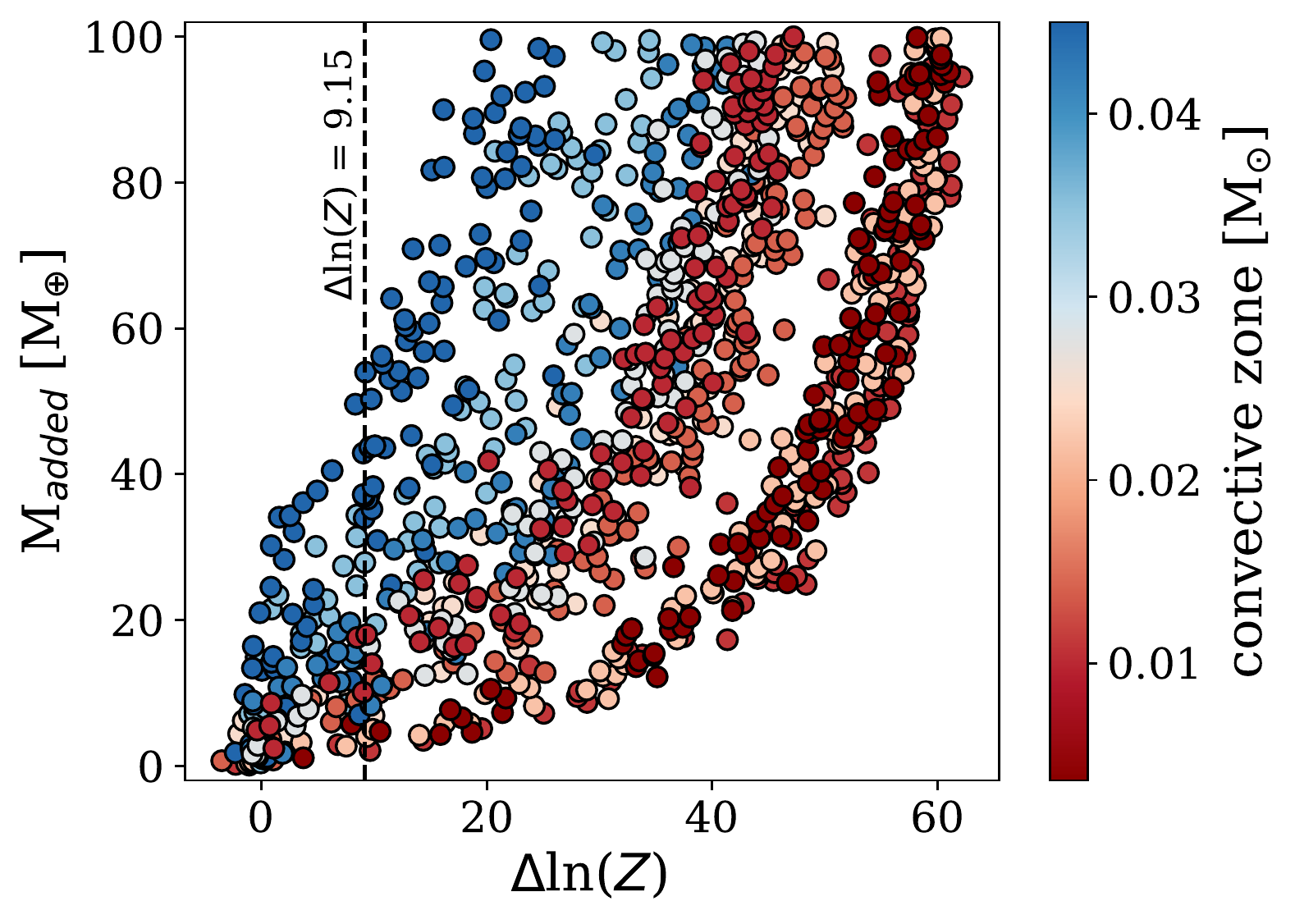} 
    \end{minipage}
\caption{The left panel displays the difference in Bayesian evidence values between the engulfment and flat models $\Delta$ln($Z$) for the eleven twin systems in our engulfment sample (black), with the two $\Delta$ln($Z$) probability density functions for simulated systems that have (red) or have not (blue) undergone engulfment. These synthetic samples are each composed of 1000 systems randomly drawn with replacement from our ten twin systems (excluding HAT-P-4). For the simulated engulfment systems, we added 10 $M_{\oplus}$ of bulk Earth composition material to the planet host star and computed abundances according to our engulfment model. The synthetic engulfment and non-engulfment distributions exhibit significant overlap. The right panel displays $\Delta$ln($Z$) values corresponding to 0.1$-$100 $M_{\oplus}$ simulated engulfment systems. The colors represent the engulfing star convective zone mass, and the maximum $\Delta$ln($Z$) value for the synthetic non-engulfment distribution of 9.15 is marked by the dashed line.}
\label{fig:figure3}
\end{figure*}

The \texttt{SME}-determined stellar parameters ($T_{\textrm{eff}}$, log$g$) for the engulfment sample are provided in Table \ref{tab:table2}. The \texttt{SME} stellar parameters are more accurate than those initially used to select our engulfment sample, and seven stars (WASP-3 C, HD 23596 B, PR0211 B, HAT-P-41 B, Kepler-410 B, WASP-70 B, Kepler-1150 B) have \texttt{SME}-determined $T_{\textrm{eff}}$ below our sample cutoff 4700 K. Thus, these systems were removed from our engulfment analysis, leaving 29 binaries in our sample that include all eleven twin systems. The \texttt{SME}-determined abundances are given in Table \ref{tab:table4}, and associated errors in Table \ref{tab:table5}. The abundance errors are estimated from two sources: the SNR of the HIRES spectra as mentioned above, and the scatter in measured abundances from different observations of the same target. To quantify how these error sources affect abundance predictions, \citet{brewer2018} ran \texttt{SME} on a set of simulated solar and cool star spectra with varying amounts of added Gaussian random noise that mimic varying SNR levels (Table 2, \citealt{brewer2018}). We conducted a similar investigation with real data using Keck-HIRES observations of eight bright stars spanning a range of $T_{\textrm{eff}}$ and [Fe/H] at five different SNR levels, and found that the scatter in \texttt{SME}-determined abundances agrees with the abundance errors reported in \citet{brewer2018}. This error analysis is described more fully in Appendix \ref{sec:snr_test}.

\subsection{Lithium Measurements} \label{sec:lithium}
Lithium abundances provide an independent line of evidence for planet engulfment. Unlike other refractory species, lithium is destroyed in thermonuclear reactions at comparatively low temperatures ($T$ $\approx$ 3 $\times$ $10^{6}$ K), making it short-lived in stellar photospheres and a potential tracer of stellar age (e.g., \citealt{berger2018}). Thus, enhanced surface lithium in stars that are not particularly young may signify recent events that modified stellar chemistry beyond birth compositions, such as planet engulfment. 

The Li I doublet at 6708 {\AA} was used to measure Li abundances for our sample of planet host binaries. First, we derived Li equivalent width (EW) measurements. This was done with spectra that were continuum-normalized through removal of the blaze function, then Doppler-corrected through cross-correlation with the rest-wavelength, National Solar Observatory solar spectrum \citep{wallace2011} as implemented in the \texttt{SpecMatch-Emp} package \citep{yee2017}. We followed the procedure outlined in \citet{berger2018} to calculate Li EWs. In brief, the \texttt{LMFIT} \citep{newville2014} Levenberg-Marquardt minimization routine implemented in Python was used to fit a four component composite model to the Li I doublet region. The components consisted of a constant to accommodate the continuum, two Gaussians for the two Li I features at 6707.76 {\AA} and 6707.91 {\AA}, and another Gaussian for the nearby Fe I feature at 6707.44 {\AA}. Only the continuum constant and two Li I Gaussians were considered in the Li EW calculation. Li EW measurement uncertainties were taken as the quadratic sum of the statistical photometric error due to SNR/pixel, and the range in EW measurements when modifying the continuum placement \citep{cayrel1988,bertran2015}. 

Li abundances were derived from the EW measurements with the \texttt{MOOG} \citep{sneden1973} spectral synthesis code. We chose \texttt{MOOG} over \texttt{SME} because the \texttt{SME} line list in our implementation from \citet{brewer2016} does not include Li spectral features. Instead, we used the \texttt{MOOG} \emph{blends} routine. \texttt{MOOG} was implemented via the Python wrappers \texttt{pymoog}\footnote{\url{https://github.com/MingjieJian/pymoog/}} and \texttt{pymoogi}\footnote{\url{https://github.com/madamow/pymoogi/}}, where \texttt{pymoog} was used to select an appropriate model atmosphere from a provided library of Kurucz ATLAS9 model grids, and the Li abundances were calculated via the \emph{blends} routine contained in \texttt{pymoogi} from Li EW measurements. In this step, the errors on stellar parameters were incorporated by simultaneously sampling from Gaussian distributions with widths equal to the uncertainties on $T_{\textrm{eff}}$, log$g$, and [Fe/H] 100 times. The scatter of the resulting abundance measurements was then added in quadrature with the difference in Li abundance from including the Li EW uncertainty discussed above. The result is our total Li abundance uncertainty. The engulfment sample Li EWs and abundances are provided in Table \ref{tab:table6}.

\begin{deluxetable*}{lrrrrrrrrrrrrrrr}
\centering
\tablewidth{0.999\textwidth}
\tabletypesize{\footnotesize}
\tablewidth{0pt}
\tablecaption{Abundances of Engulfment Sample Stars \label{tab:table4}}
\tablecolumns{16}
\tablehead{
\colhead{Name} &
\colhead{[C/H]} &
\colhead{[N/H]} &
\colhead{[O/H]} &
\colhead{[Na/H]} &
\colhead{[Mn/H]} &
\colhead{[Cr/H]} &
\colhead{[Si/H]} &
\colhead{[Fe/H]} &
\colhead{[Mg/H]} &
\colhead{[Ni/H]} &
\colhead{[V/H]} &
\colhead{[Ca/H]} &
\colhead{[Ti/H]} &
\colhead{[Al/H]} &
\colhead{[Y/H]} \\
\colhead{} &
\colhead{dex} &
\colhead{dex} &
\colhead{dex} &
\colhead{dex} &
\colhead{dex} &
\colhead{dex} &
\colhead{dex} &
\colhead{dex} &
\colhead{dex} &
\colhead{dex} &
\colhead{dex} &
\colhead{dex} &
\colhead{dex} &
\colhead{dex} &
\colhead{dex} 
}
\startdata
 HAT-P-4 A* &   0.17 &   0.31 &   0.27 &    0.21 &    0.23 &    0.30 &    0.25 &    0.33 &    0.28 &    0.21 &   0.28 &    0.24 &    0.29 &    0.30 &   0.39 \\
      HAT-P-4 B &   0.12 &   0.23 &   0.19 &    0.15 &    0.16 &    0.17 &    0.19 &    0.20 &    0.16 &    0.10 &   0.17 &    0.19 &    0.19 &    0.19 &   0.22 \\
    HD 132563 A &  $-$0.11 &   0.18 &   0.07 &   $-$0.19 &   $-$0.31 &   $-$0.12 &   $-$0.10 &   $-$0.11 &   $-$0.13 &   $-$0.20 &  $-$0.14 &   $-$0.09 &   $-$0.07 &   $-$0.26 &  $-$0.10 \\
   HD 132563 B* &  $-$0.10 &  $-$0.09 &  $-$0.02 &   $-$0.19 &   $-$0.28 &   $-$0.13 &   $-$0.11 &   $-$0.12 &   $-$0.12 &   $-$0.19 &  $-$0.15 &   $-$0.09 &   $-$0.09 &   $-$0.25 &  $-$0.10 \\
   HD 133131 A* &  $-$0.19 &  $-$0.33 &  $-$0.17 &   $-$0.26 &   $-$0.22 &   $-$0.22 &   $-$0.21 &   $-$0.23 &   $-$0.19 &   $-$0.20 &  $-$0.28 &   $-$0.37 &   $-$0.26 &   $-$0.27 &  $-$0.32 \\
   HD 133131 B* &  $-$0.19 &  $-$0.25 &  $-$0.15 &   $-$0.27 &   $-$0.22 &   $-$0.23 &   $-$0.22 &   $-$0.24 &   $-$0.20 &   $-$0.24 &  $-$0.29 &   $-$0.39 &   $-$0.27 &   $-$0.29 &  $-$0.33 \\
   $\omega$ Ser A* &  $-$0.24 &   0.17 &  $-$0.14 &    0.00 &    0.15 &    0.04 &   $-$0.18 &    0.13 &   $-$0.07 &    0.14 &  $-$0.03 &    0.08 &    0.09 &    0.00 &   0.41 \\
    $\omega$ Ser B &  $-$0.16 &  $-$0.31 &  $-$0.07 &   $-$0.21 &   $-$0.25 &   $-$0.17 &   $-$0.14 &   $-$0.18 &   $-$0.14 &   $-$0.21 &  $-$0.15 &   $-$0.13 &   $-$0.13 &   $-$0.17 &  $-$0.13 \\
   HD 178911 A &   0.13 &   0.34 &   0.23 &    0.31 &    0.23 &    0.20 &    0.15 &    0.20 &    0.12 &    0.20 &   0.28 &    0.25 &    0.20 &    0.23 &   0.17 \\
   HD 178911 B* &   0.18 &   0.24 &   0.17 &    0.28 &    0.26 &    0.21 &    0.22 &    0.21 &    0.18 &    0.24 &   0.21 &    0.21 &    0.20 &    0.22 &   0.14 \\
     16 Cyg A &   0.07 &   0.05 &   0.13 &    0.10 &    0.08 &    0.08 &    0.08 &    0.09 &    0.08 &    0.10 &   0.11 &    0.10 &    0.09 &    0.12 &   0.03 \\
      16 Cyg B* &   0.04 &   0.05 &   0.06 &    0.08 &    0.05 &    0.06 &    0.06 &    0.06 &    0.06 &    0.07 &   0.06 &    0.07 &    0.08 &    0.09 &   0.03 \\
   HD 202772 A* &   0.15 &   0.55 &   0.47 &    0.31 &    0.24 &    0.37 &    0.27 &    0.35 &    0.19 &    0.31 &   0.20 &    0.41 &    0.38 &    0.25 &   0.57 \\
    HD 202772 B &   0.16 &   0.45 &  $-$0.01 &    0.29 &    0.33 &    0.26 &    0.25 &    0.29 &    0.17 &    0.28 &   0.17 &    0.32 &    0.25 &    0.27 &   0.33 \\
    HAT-P-1 A &   0.07 &   0.13 &   0.21 &    0.13 &    0.05 &    0.14 &    0.13 &    0.15 &    0.09 &    0.11 &   0.12 &    0.21 &    0.15 &    0.11 &   0.17 \\
     HAT-P-1 B* &   0.11 &   0.15 &   0.17 &    0.12 &    0.14 &    0.16 &    0.16 &    0.16 &    0.14 &    0.16 &   0.14 &    0.18 &    0.15 &    0.15 &   0.24 \\
   Kepler-25 A* &  $-$0.06 &   0.06 &   0.16 &   $-$0.07 &   $-$0.14 &    0.00 &   $-$0.01 &    0.01 &   $-$0.05 &   $-$0.09 &  $-$0.08 &    0.05 &    0.05 &   $-$0.19 &  $-$0.03 \\
    Kepler-25 B &  $-$0.01 &  $-$0.33 &  $-$0.07 &   $-$0.03 &    0.01 &    0.04 &    0.03 &    0.03 &   $-$0.04 &   $-$0.02 &   0.00 &    0.08 &    0.00 &    0.00 &  $-$0.08 \\
     WASP-94 A* &   0.21 &   0.39 &   0.32 &    0.34 &    0.37 &    0.34 &    0.28 &    0.34 &    0.28 &    0.37 &   0.22 &    0.35 &    0.34 &    0.26 &   0.44 \\
      WASP-94 B &   0.24 &   0.35 &   0.34 &    0.37 &    0.43 &    0.30 &    0.30 &    0.32 &    0.29 &    0.35 &   0.20 &    0.38 &    0.32 &    0.34 &   0.35 \\
      HD 20781* &  $-$0.06 &  $-$0.16 &   0.03 &   $-$0.15 &   $-$0.11 &   $-$0.05 &   $-$0.07 &   $-$0.05 &   $-$0.04 &   $-$0.10 &  $-$0.05 &   $-$0.04 &   $-$0.03 &   $-$0.05 &  $-$0.18 \\
      HD 20782* &  $-$0.06 &  $-$0.13 &  $-$0.01 &   $-$0.17 &   $-$0.16 &   $-$0.07 &   $-$0.08 &   $-$0.06 &   $-$0.06 &   $-$0.11 &  $-$0.08 &   $-$0.05 &   $-$0.05 &   $-$0.06 &  $-$0.14 \\
    HD 40979 A* &   0.17 &   0.37 &   0.29 &    0.29 &    0.29 &    0.27 &    0.25 &    0.27 &    0.21 &    0.25 &   0.24 &    0.29 &    0.23 &    0.18 &   0.23 \\
     HD 40979 B &   0.24 &   0.13 &   0.19 &    0.31 &    0.27 &    0.22 &    0.25 &    0.27 &    0.19 &    0.26 &   0.23 &    0.30 &    0.24 &    0.25 &   0.05 \\
      KELT-2 A* &   0.05 &   0.36 &   0.38 &    0.19 &    0.08 &    0.18 &    0.15 &    0.20 &    0.07 &    0.14 &   0.08 &    0.22 &    0.16 &    0.07 &   0.18 \\
       KELT-2 B &   0.20 &  $-$0.13 &   0.10 &    0.30 &    0.20 &    0.06 &    0.36 &    0.21 &    0.07 &    0.16 &   0.12 &    0.25 &    0.10 &    0.36 &  -0.14 \\
    WASP-173 A* &   0.15 &   0.28 &   0.25 &    0.26 &    0.30 &    0.24 &    0.18 &    0.23 &    0.16 &    0.22 &   0.17 &    0.27 &    0.26 &    0.21 &   0.22 \\
     WASP-173 B &   0.07 &   0.18 &   0.03 &    0.16 &    0.19 &    0.16 &    0.15 &    0.17 &    0.14 &    0.18 &   0.15 &    0.16 &    0.19 &    0.17 &   0.14 \\
    WASP-180 A* &  $-$0.09 &   0.31 &   0.13 &   $-$0.08 &   $-$0.02 &    0.08 &    0.06 &    0.10 &    0.00 &   $-$0.04 &  $-$0.20 &    0.17 &    0.07 &   $-$0.37 &   0.18 \\
     WASP-180 B &  $-$0.06 &  $-$0.02 &   0.15 &   $-$0.08 &   $-$0.12 &    0.02 &   $-$0.02 &    0.01 &   $-$0.07 &   $-$0.12 &  $-$0.07 &    0.09 &    0.06 &   $-$0.09 &   0.06 \\
  Kepler-515 A* &  $-$0.14 &  $-$0.22 &  $-$0.07 &   $-$0.23 &   $-$0.19 &   $-$0.17 &   $-$0.11 &   $-$0.16 &   $-$0.10 &   $-$0.18 &  $-$0.09 &   $-$0.16 &   $-$0.09 &   $-$0.16 &  $-$0.19 \\
   Kepler-515 B &   0.04 &  $-$0.20 &   0.13 &   $-$0.16 &   $-$0.26 &   $-$0.18 &   $-$0.03 &   -0.20 &   $-$0.13 &   $-$0.17 &  $-$0.12 &   $-$0.09 &   $-$0.11 &   $-$0.08 &  $-$0.43 \\
   Kepler-477 A &  $-$0.15 &  $-$0.49 &   0.07 &   $-$0.47 &   $-$0.53 &   $-$0.40 &   $-$0.27 &   $-$0.39 &   $-$0.34 &   $-$0.43 &  $-$0.28 &   $-$0.33 &   $-$0.26 &   $-$0.29 &  $-$0.65 \\
  Kepler-477 B* &  $-$0.28 &  $-$0.56 &  $-$0.11 &   $-$0.50 &   $-$0.57 &   $-$0.42 &   $-$0.37 &   $-$0.44 &   $-$0.34 &   $-$0.43 &  $-$0.36 &   $-$0.35 &   $-$0.33 &   $-$0.38 &  $-$0.56 \\
 Kepler-1063 A* &   0.14 &   0.16 &   0.09 &    0.24 &    0.20 &    0.17 &    0.15 &    0.16 &    0.14 &    0.19 &   0.12 &    0.21 &    0.14 &    0.18 &   0.10 \\
  Kepler-1063 B &   0.20 &   0.18 &   0.20 &    0.28 &    0.28 &    0.24 &    0.22 &    0.23 &    0.19 &    0.25 &   0.21 &    0.27 &    0.23 &    0.27 &   0.22 \\
      WASP-3 A* &  $-$0.14 &   0.40 &   0.12 &   $-$0.21 &   $-$0.24 &   $-$0.02 &   $-$0.01 &   $-$0.01 &   $-$0.09 &   $-$0.14 &  $-$0.05 &    0.01 &    0.01 &   $-$0.41 &  $-$0.09 \\
       WASP-3 C &  $-$0.11 &  $-$0.61 &  $-$0.27 &   $-$0.02 &   $-$0.14 &   $-$0.12 &    0.05 &   $-$0.07 &   $-$0.13 &   $-$0.08 &  $-$0.09 &    0.05 &   $-$0.07 &    0.02 &  $-$0.08 \\
    WASP-160 A &   0.15 &   0.19 &   0.23 &    0.18 &    0.15 &    0.17 &    0.14 &    0.15 &    0.15 &    0.14 &   0.15 &    0.17 &    0.14 &    0.16 &   0.10 \\
    WASP-160 B* &   0.15 &   0.25 &   0.16 &    0.24 &    0.19 &    0.21 &    0.18 &    0.17 &    0.19 &    0.22 &   0.16 &    0.19 &    0.19 &    0.21 &   0.17 \\
      HD 80606* &   0.26 &   0.39 &   0.26 &    0.44 &    0.40 &    0.28 &    0.31 &    0.30 &    0.28 &    0.34 &   0.28 &    0.27 &    0.27 &    0.33 &   0.20 \\
       HD 80607 &   0.26 &   0.39 &   0.24 &    0.46 &    0.38 &    0.28 &    0.30 &    0.30 &    0.25 &    0.34 &   0.27 &    0.29 &    0.27 &    0.34 &   0.15 \\
         XO-2N* &   0.35 &   0.43 &   0.34 &    0.46 &    0.43 &    0.40 &    0.36 &    0.42 &    0.35 &    0.43 &   0.38 &    0.40 &    0.38 &    0.44 &   0.26 \\
         XO-2S* &   0.31 &   0.43 &   0.31 &    0.44 &    0.29 &    0.36 &    0.29 &    0.33 &    0.30 &    0.29 &   0.34 &    0.37 &    0.34 &    0.35 &   0.18 \\
    HD 99491 &   0.23 &   0.34 &   0.24 &    0.38 &    0.31 &    0.28 &    0.27 &    0.28 &    0.24 &    0.29 &   0.26 &    0.26 &    0.25 &    0.32 &   0.18 \\
      HD 99492* &   0.31 &   0.37 &   0.26 &    0.50 &    0.36 &    0.29 &    0.28 &    0.32 &    0.29 &    0.34 &   0.28 &    0.32 &    0.28 &    0.37 &   0.14 \\
   HD 106515 A* &   0.11 &   0.13 &   0.22 &    0.08 &    0.02 &    0.05 &    0.08 &    0.05 &    0.14 &    0.07 &   0.10 &    0.08 &    0.14 &    0.16 &  $-$0.07 \\
    HD 106515 B &   0.15 &   0.12 &   0.26 &    0.09 &    0.04 &    0.07 &    0.10 &    0.07 &    0.13 &    0.07 &   0.12 &    0.09 &    0.15 &    0.18 &  $-$0.06 \\
     WASP-64 A &   0.11 &   0.11 &   0.16 &    0.18 &    0.17 &    0.13 &    0.13 &    0.14 &    0.14 &    0.16 &   0.11 &    0.15 &    0.15 &    0.17 &   0.08 \\
     WASP-64 B* &   0.11 &   0.15 &   0.08 &    0.20 &    0.18 &    0.16 &    0.15 &    0.16 &    0.16 &    0.17 &   0.16 &    0.15 &    0.16 &    0.18 &   0.05 \\
    WASP-127 A* &  $-$0.12 &  $-$0.26 &   0.00 &   $-$0.27 &   $-$0.39 &   $-$0.18 &   $-$0.16 &   $-$0.17 &   $-$0.13 &   $-$0.22 &  $-$0.14 &   $-$0.11 &   $-$0.05 &   $-$0.16 &  $-$0.23 \\
     WASP-127 B &  $-$0.19 &  $-$0.33 &  $-$0.03 &   $-$0.28 &   $-$0.36 &   $-$0.25 &   $-$0.19 &   $-$0.21 &   $-$0.15 &   $-$0.25 &  $-$0.15 &   $-$0.17 &   $-$0.14 &   $-$0.20 &  $-$0.31 
\enddata
\tablecomments{This is a subset of a table that lists the \texttt{SME}-determined elemental abundances for stars in the engulfment sample. $T_{\textrm{eff}}$ and log$g$ were calculated by applying \texttt{SME} to the Keck-HIRES spectra. The brighter component of each binary pair is denoted as $`$A', and the fainter component as $`$B'. The planet hosts are marked with *.
\vspace{1mm}
\newline (This table is available in its entirety in machine-readable form.)}
\end{deluxetable*}

\begin{deluxetable*}{llrrrrrrrrrrrrrrr}
\centering
\tablewidth{0.999\textwidth}
\tabletypesize{\footnotesize}
\tablewidth{0pt}
\tablecaption{Abundance Errors of Engulfment Sample Stars \label{tab:table5}}
\tablecolumns{14}
\tablehead{
\colhead{Name} &
\colhead{SNR/pix} &
\colhead{$\sigma$[C/H]} &
\colhead{$\sigma$[N/H]} &
\colhead{$\sigma$[O/H]} &
\colhead{$\sigma$[Na/H]} &
\colhead{$\sigma$[Mn/H]} &
\colhead{$\sigma$[Cr/H]} &
\colhead{$\sigma$[Si/H]} &
\colhead{$\sigma$[Fe/H]} &
\colhead{$\sigma$[Mg/H]} &
\colhead{$\sigma$[Ni/H]} &
\colhead{$\sigma$[V/H]} &
\colhead{...} \\
\colhead{} &
\colhead{} &
\colhead{dex} &
\colhead{dex} &
\colhead{dex} &
\colhead{dex} &
\colhead{dex} &
\colhead{dex} &
\colhead{dex} &
\colhead{dex} &
\colhead{dex} &
\colhead{dex} &
\colhead{dex} &
\colhead{}
}
\startdata
   HAT-P-4 A* &  150 &  0.011 &  0.042 &  0.019 &  0.014 &  0.010 &  0.007 &  0.009 &  0.006 &  0.008 &  0.008 &  0.017 &  ... \\
    HAT-P-4 B &  139 &  0.011 &  0.042 &  0.019 &  0.014 &  0.010 &  0.007 &  0.009 &  0.006 &  0.008 &  0.008 &  0.017 &  ... \\
  HD 132563 A &  200 &  0.011 &  0.042 &  0.019 &  0.014 &  0.010 &  0.007 &  0.009 &  0.006 &  0.008 &  0.008 &  0.017 &  ... \\
 HD 132563 B* &  200 &  0.011 &  0.042 &  0.019 &  0.014 &  0.010 &  0.007 &  0.009 &  0.006 &  0.008 &  0.008 &  0.017 &  ... \\
 HD 133131 A* &  201 &  0.011 &  0.042 &  0.019 &  0.014 &  0.010 &  0.007 &  0.009 &  0.006 &  0.008 &  0.008 &  0.017 &  ... \\
 HD 133131 B* &  200 &  0.011 &  0.042 &  0.019 &  0.014 &  0.010 &  0.007 &  0.009 &  0.006 &  0.008 &  0.008 &  0.017 &  ... \\
 $\omega$ Ser A* &  253 &  0.011 &  0.042 &  0.019 &  0.014 &  0.010 &  0.007 &  0.009 &  0.006 &  0.008 &  0.008 &  0.017 &  ... \\
  $\omega$ Ser B &  200 &  0.011 &  0.042 &  0.019 &  0.014 &  0.010 &  0.007 &  0.009 &  0.006 &  0.008 &  0.008 &  0.017 &  ... \\
  HD 178911 A &  202 &  0.011 &  0.042 &  0.019 &  0.014 &  0.010 &  0.007 &  0.009 &  0.006 &  0.008 &  0.008 &  0.017 &  ... \\
 HD 178911 B* &  253 &  0.011 &  0.042 &  0.019 &  0.014 &  0.010 &  0.007 &  0.009 &  0.006 &  0.008 &  0.008 &  0.017 &  ... \\
     16 Cyg A &  205 &  0.011 &  0.042 &  0.019 &  0.014 &  0.010 &  0.007 &  0.009 &  0.006 &  0.008 &  0.008 &  0.017 &  ... \\
    16 Cyg B* &  200 &  0.011 &  0.042 &  0.019 &  0.014 &  0.010 &  0.007 &  0.009 &  0.006 &  0.008 &  0.008 &  0.017 &  ... \\
 HD 202772 A* &  141 &  0.011 &  0.042 &  0.019 &  0.014 &  0.010 &  0.007 &  0.009 &  0.006 &  0.008 &  0.008 &  0.017 &  ... \\
  HD 202772 B &  141 &  0.011 &  0.042 &  0.019 &  0.014 &  0.010 &  0.007 &  0.009 &  0.006 &  0.008 &  0.008 &  0.017 &  ... \\
    HAT-P-1 A &  219 &  0.011 &  0.042 &  0.019 &  0.014 &  0.010 &  0.007 &  0.009 &  0.006 &  0.008 &  0.008 &  0.017 &  ... \\
   HAT-P-1 B* &  98 &  0.011 &  0.042 &  0.019 &  0.014 &  0.010 &  0.007 &  0.009 &  0.006 &  0.008 &  0.008 &  0.017 &  ... \\
 Kepler-25 A* &  167 &  0.011 &  0.042 &  0.019 &  0.014 &  0.010 &  0.007 &  0.009 &  0.006 &  0.008 &  0.008 &  0.017 &  ... \\
  Kepler-25 B &   40 &  0.014 &  0.082 &  0.035 &  0.021 &  0.016 &  0.014 &  0.016 &  0.008 &  0.013 &  0.012 &  0.031 &  ... \\
   WASP-94 A* &   58 &  0.013 &  0.069 &  0.028 &  0.020 &  0.014 &  0.010 &  0.014 &  0.008 &  0.012 &  0.010 &  0.027 &  ... \\
    WASP-94 B &   57 &  0.013 &  0.070 &  0.029 &  0.020 &  0.014 &  0.010 &  0.014 &  0.008 &  0.012 &  0.010 &  0.027 &  ... \\
    HD 20781* &  200 &  0.011 &  0.042 &  0.019 &  0.014 &  0.010 &  0.007 &  0.009 &  0.006 &  0.008 &  0.008 &  0.017 &  ... \\
    HD 20782* &  202 &  0.011 &  0.042 &  0.019 &  0.014 &  0.010 &  0.007 &  0.009 &  0.006 &  0.008 &  0.008 &  0.017 &  ... \\
  HD 40979 A* &  253 &  0.011 &  0.042 &  0.019 &  0.014 &  0.010 &  0.007 &  0.009 &  0.006 &  0.008 &  0.008 &  0.017 &  ... \\
   HD 40979 B &  141 &  0.011 &  0.042 &  0.019 &  0.014 &  0.010 &  0.007 &  0.009 &  0.006 &  0.008 &  0.008 &  0.017 &  ... \\
    KELT-2 A* &  142 &  0.011 &  0.042 &  0.019 &  0.014 &  0.010 &  0.007 &  0.009 &  0.006 &  0.008 &  0.008 &  0.017 &  ... \\
     KELT-2 B &   51 &  0.013 &  0.068 &  0.028 &  0.021 &  0.015 &  0.011 &  0.013 &  0.007 &  0.012 &  0.009 &  0.026 &  ... \\
  WASP-173 A* &   51 &  0.013 &  0.072 &  0.031 &  0.019 &  0.014 &  0.010 &  0.014 &  0.008 &  0.013 &  0.011 &  0.027 &  ... \\
   WASP-173 B &   51 &  0.013 &  0.071 &  0.030 &  0.020 &  0.014 &  0.010 &  0.014 &  0.008 &  0.013 &  0.010 &  0.027 &  ... \\
  WASP-180 A* &   62 &  0.013 &  0.066 &  0.026 &  0.022 &  0.015 &  0.011 &  0.013 &  0.007 &  0.011 &  0.008 &  0.026 &  ... \\
   WASP-180 B &   40 &  0.016 &  0.092 &  0.036 &  0.025 &  0.020 &  0.019 &  0.020 &  0.009 &  0.011 &  0.012 &  0.037 &  ... \\
Kepler-515 A* &   49 &  0.013 &  0.073 &  0.032 &  0.019 &  0.014 &  0.010 &  0.014 &  0.008 &  0.013 &  0.011 &  0.027 &  ... \\
 Kepler-515 B &   51 &  0.013 &  0.070 &  0.030 &  0.020 &  0.014 &  0.010 &  0.014 &  0.008 &  0.012 &  0.010 &  0.027 &  ... \\
 Kepler-477 A &   40 &  0.017 &  0.097 &  0.037 &  0.026 &  0.021 &  0.019 &  0.020 &  0.010 &  0.012 &  0.013 &  0.038 &  ... \\
Kepler-477 B* &   40 &  0.017 &  0.097 &  0.037 &  0.026 &  0.021 &  0.019 &  0.020 &  0.010 &  0.012 &  0.013 &  0.038 &  ... \\
Kepler-1063 A* &   51 &  0.013 &  0.073 &  0.032 &  0.019 &  0.014 &  0.010 &  0.014 &  0.008 &  0.013 &  0.011 &  0.027 &  ... \\
Kepler-1063 B &   51 &  0.013 &  0.073 &  0.032 &  0.019 &  0.014 &  0.010 &  0.014 &  0.008 &  0.013 &  0.011 &  0.027 &  ... \\
    WASP-3 A* &  170 &  0.011 &  0.042 &  0.019 &  0.014 &  0.010 &  0.007 &  0.009 &  0.006 &  0.008 &  0.008 &  0.017 &  ... \\
     WASP-3 C &   40 &  0.014 &  0.083 &  0.035 &  0.022 &  0.017 &  0.014 &  0.017 &  0.008 &  0.012 &  0.012 &  0.032 &  ... \\
   WASP-160 A &  51 &  0.013 &  0.071 &  0.031 &  0.020 &  0.014 &  0.010 &  0.014 &  0.008 &  0.013 &  0.011 & 0.027 & ... \\
  WASP-160 B* &  51 &  0.013 &  0.071 &  0.030 &  0.020 &  0.014 &  0.010 &  0.014 &  0.008 &  0.013 &  0.010 &  0.027 &  ... \\
    HD 80606* &  201 &  0.011 &  0.042 &  0.019 &  0.014 &  0.010 &  0.007 &  0.009 &  0.006 &  0.008 &  0.008 &  0.017 &  ... \\
     HD 80607 &  200 &  0.011 &  0.042 &  0.019 &  0.014 &  0.010 &  0.007 &  0.009 &  0.006 &  0.008 &  0.008 &  0.017 &  ... \\
       XO-2 N* &  200 &  0.011 &  0.042 &  0.019 &  0.014 &  0.010 &  0.007 &  0.009 &  0.006 &  0.008 &  0.008 &  0.017 &  ... \\
       XO-2 S* &  141 &  0.011 &  0.042 &  0.019 &  0.014 &  0.010 &  0.007 &  0.009 &  0.006 &  0.008 &  0.008 & 0.017 & ... \\
     HD 99491 &  200 &  0.011 &  0.042 &  0.019 &  0.014 &  0.010 &  0.007 &  0.009 &  0.006 &  0.008 &  0.008 &  0.017 &  ... \\
    HD 99492* &  200 &  0.011 &  0.042 &  0.019 &  0.014 &  0.010 &  0.007 &  0.009 &  0.006 &  0.008 &  0.008 &  0.017 &  ... \\
 HD 106515 A* &  200 &  0.011 &  0.042 &  0.019 &  0.014 &  0.010 &  0.007 &  0.009 &  0.006 &  0.008 &  0.008 &  0.017 &  ... \\
  HD 106515 B &  200 &  0.011 &  0.042 &  0.019 &  0.014 &  0.010 &  0.007 &  0.009 &  0.006 &  0.008 &  0.008 &  0.017 &  ... \\
    WASP-64 A &  98 &  0.011 &  0.042 &  0.019 &  0.014 &  0.010 &  0.007 &  0.009 &  0.006 &  0.008 &  0.008 &  0.017 &  ... \\
   WASP-64 B* &   63 &  0.013 &  0.063 &  0.023 &  0.023 &  0.015 &  0.011 &  0.013 &  0.007 &  0.010 &  0.007 &  0.026 &  ... \\
  WASP-127 A* &  200 &  0.011 &  0.042 &  0.019 &  0.014 &  0.010 &  0.007 &  0.009 &  0.006 &  0.008 &  0.008 &  0.017 &  ... \\
   WASP-127 B &   69 &  0.012 &  0.057 &  0.022 &  0.020 &  0.014 &  0.010 &  0.012 &  0.007 &  0.009 &  0.007 &  0.023 &  ... 
\enddata
\tablecomments{This is a subset of a table that lists the \texttt{SME}-determined elemental abundance errors for stars in the engulfment sample. The brighter component of each binary pair is denoted as $`$A', and the fainter component as $`$B'. The planet hosts are marked with *. 
\vspace{1mm}
\newline (This table is available in its entirety in machine-readable form.)}
\end{deluxetable*}

\begin{deluxetable}{lrrr}
\centering
\tablewidth{0.5\textwidth}
\tabletypesize{\footnotesize}
\tablewidth{0pt}
\setlength{\tabcolsep}{8pt}
\tablecaption{Lithium Measurements \label{tab:table6}}
\tablecolumns{4}
\tablehead{
\colhead{Name} &
\colhead{$EW_{\textrm{Li}}$} &
\colhead{A(Li)} &
\colhead{$\Delta$A(Li)} \\
\colhead{} &
\colhead{m{\AA}} &
\colhead{dex} &
\colhead{dex}
}
\startdata
      HD 23596 A* &  73.18 $\pm$ 2.55 &  2.68 $\pm$ 0.03 &   2.83 $\pm$ 0.08 \\
     HD 23596 B &  17.53 $\pm$ 2.69 & $-$0.14 $\pm$ 0.07 &  -- \\
        WASP-3 A* &  18.27 $\pm$ 1.06 &  2.26 $\pm$ 0.03 &        $>$ 2.83 \\
        WASP-3 C &   2.16 $\pm$ 4.06 & $<$ $-$0.57 &       --  \\
       KELT-4 A* &  24.64 $\pm$ 1.04 &  2.39 $\pm$ 0.03 &        2.68 $\pm$ 0.18  \\
        KELT-4 B &   2.03 $\pm$ 1.01 & $-$0.29 $\pm$ 0.18 &     --\\
    Kepler-410 A* &  13.98 $\pm$ 2.08 &  2.12 $\pm$ 0.07 &       $>$ 2.50 \\
   Kepler-410 B &   0.00 $\pm$ 1.36 & $<$ $-$0.38 &       -- \\
     Kepler-25 A* &  23.69 $\pm$ 1.07 &  2.31 $\pm$ 0.03 &       $>$ 2.32 \\
     Kepler-25 B &   3.24 $\pm$ 3.58 & $<$ $-$0.02  &       --  \\
      HD 40979 A* &  79.33 $\pm$ 0.61 &  2.86 $\pm$ 0.02 &        2.24 $\pm$ 0.06 \\
     HD 40979 B &  11.30 $\pm$ 1.12 &  0.62 $\pm$ 0.05 &       -- \\
    Kepler-104 A* &  17.07 $\pm$ 0.82 &  1.84 $\pm$ 0.03 &        $>$ 1.87 \\
   Kepler-104 B &   0.00 $\pm$ 0.70 & $<$ $-$0.03 &       --  \\
   WASP-70 A* &   3.29 $\pm$ 2.85 &  1.03 $\pm$ 0.28 &       $>$ 1.71 \\
    WASP-70 B &   1.20 $\pm$ 2.65 & $<$ $-$0.68 &       -- \\
      HAT-P-41 A* &   1.44 $\pm$ 0.76 &  1.22 $\pm$ 0.19 &        $>$ 1.70 $\pm$ 0.20 \\
      HAT-P-41 B &   0.00 $\pm$ 2.49 & $<$ $-$0.48 &       -- \\
       WASP-127 A* &  27.13 $\pm$ 1.56 &  2.03 $\pm$ 0.03 &        $>$ 1.60 \\
      WASP-127 B &   0.00 $\pm$ 2.23 &  $<$ 0.43 &       -- \\
     WASP-173 A* &   0.00 $\pm$ 2.70 &  $<$ 1.53 &       $<$ 1.37 \\
      WASP-173 B &   0.94 $\pm$ 2.67 &  $<$ 0.16 &       --  \\
     Kepler-99 B* &  0.00 $\pm$ 2.78 &  $<$ 0.15 &       $<$ 1.07 \\
      Kepler-99 A &  0.48 $\pm$ 1.18 & $<$ $-$0.92 &    -- \\
      WASP-94 A* &   9.73 $\pm$ 3.33 &  1.75 $\pm$ 0.13 &  $>$ 1.03 \\
       WASP-94 B &   1.07 $\pm$ 3.11 &  $<$ 0.72  &    -- 
\enddata
\tablecomments{This is a subset of a table that lists the $EW_{\textrm{Li}}$ and A(Li) measurements for stars in the engulfment sample, ranked by their $\Delta$A(Li). In cases where the Li EW is smaller than the associated uncertainty, A(Li) is reported as an upper limit. The brighter component of each binary pair is denoted as $`$A', and the fainter component as $`$B'. The planet hosts are marked with *.  
\vspace{1mm}
\newline (This table is available in its entirety in machine-readable form.)}
\end{deluxetable}

\section{Engulfment Model} \label{sec:model}
We present a framework similar to that of \citet{oh2018} for estimating the remaining mass of bulk Earth composition \citep{mcdonough2003} material engulfed in one star given abundance measurements for a binary pair. We emphasize \emph{remaining} here because the initial refractory enrichment in stellar photospheres following engulfment is depleted over time; once the system is observed, there will be less refractory material in the engulfing star photosphere than was immediately present after the engulfment event (see Section \ref{sec:timescales} for our analysis of engulfment signature timescales).

From the stellar abundances of the engulfing star [X/H], we can express the mass fraction of each element X as:

\begin{eqnarray}
f_{\textrm{X,photo}} = \frac{10^{[\textrm{X/H}]} m_{\textrm{X}}}{\Sigma_{\textrm{X}} 10^{[\textrm{X/H}]} m_{\textrm{X}}} \hspace{0.5mm},
\end{eqnarray}

\noindent where $m_{\textrm{X}}$ is the mass of each element in atomic mass units. We note that this approach of computing mass fraction rather than number density fraction should be appropriate for our systems, namely binaries composed of stars with low $Z$. Assuming a total mass of accreted material $M_{\textrm{acc}}$ and accreted mass fractions for each element $f_{\textrm{X,acc}}$, the abundance difference is

\begin{eqnarray}
\Delta[\textrm{X/H}] = \textrm{log}_{10} \frac{f_{\textrm{X,photo}} \hspace{0.4mm} f_{cz} \hspace{0.4mm} M_{*} \hspace{0.4mm}+ \hspace{0.4mm} f_{\textrm{X,acc}} \hspace{0.4mm} M_{\textrm{acc}}}{f_{\textrm{X,photo}} \hspace{0.4mm} f_{cz}\hspace{0.4mm} M_{*}} \hspace{0.5mm},
\end{eqnarray}

\noindent where $f_{cz}$ is the mass fraction of the stellar convective zone. Similar calculations have been performed by, e.g., \citet{chambers2010} and \citet{mack2014,mack2016}. For more details on the engulfment model, see \citet{oh2018}. Because the modeled amount of polluting material derived from refractory enhancements depends on the convective zone mass $M_{cz}$, we adjusted $M_{cz}$ to the stellar type of the engulfing star according to the $T_{\textrm{eff}}$-$M_{cz}$ relation in \citet{pinsonneault2001}. We tested our model by applying it to the reported abundances of the Kronos-Krios system, which were also derived from Keck-HIRES spectra and \texttt{SME} \citep{brewer2016}. The model recovered 13.68\hspace{0.5mm}$\pm$\hspace{0.5mm}1.93 $M_{\oplus}$ of bulk Earth composition engulfed mass (Figure \ref{fig:figure2}), in good agreement with the reported engulfed mass of $\sim$15 $M_{\oplus}$ from \citet{oh2018}. 

Our engulfment model employs the \texttt{dynesty} nested sampling code \citep{speagle2020} to determine the Bayesian evidence for the engulfment model or a flat model of differential abundances as a function of $T_c$, shown in Figure \ref{fig:figure2} as the long-dash line. The flat model represents the case of no engulfment. We found that the engulfment model is preferred over the flat model for the Kronos-Krios system with a Bayesian evidence difference of $\Delta$ln($Z$) = 15.8.

\subsection{Bayesian Evidence} \label{sec:lnZ_criteria}
To determine the Bayesian evidence difference $\Delta$ln($Z$) that indicates a strong engulfment detection, we compared samples of simulated engulfment and non-engulfment systems. The synthetic engulfment sample was constructed by randomly drawing 1000 systems from our twin binary systems. We drew from ten of our eleven twin systems because we excluded HAT-P-4 given its potential engulfment status (see Section \ref{sec:potential_detections}). We took the planet host abundances for both stars to begin with as $\Delta$[X/H] = 0 across all elements, the added 10 $M_{\oplus}$ of bulk Earth composition material into the convective zone of the planet host star. Intrinsic scatter was then added to the abundances of the companion star according to the observed abundance scatter of 20 chemically homogeneous ($\Delta$[Fe/H] $<$ 0.05 dex) wide binaries reported in \citet{hawkins2020} (0.067 dex, 0.05 dex, 0.052 dex, 0.029 dex, 0.039 dex, 0.03 dex, 0.11 dex, 0.046 dex, 0.12 dex, 0.05 dex, 0.06 dex, 0.044 dex, 0.091 dex for C, Na, Mn, Cr, Si, Fe, Mg, Ni, V, Ca, Ti, Al, Y, respectively). Abundances for N and O were not provided in \citet{hawkins2020}, so we instead used the M67 open cluster scatter reported for these elements (0.015 dex and 0.022 dex for N and O, respectively, \citealt{bovy2016}). Further scatter was added to the companion star abundances as a function of SNR according to \citet{brewer2018} to mimic observations. The simulated non-engulfment systems were constructed by again randomly drawing 1000 systems from the twin binaries, but again excluding HAT-P-4 given its potential engulfment status. The abundances of the stars were not modified at all because we assumed that these real observations correspond to non-engulfment systems, but we again included scatter according to the 20 chemically homogeneous \citet{hawkins2020} wide binaries. 
We randomly chose the direction between the two companions when computing the differential abundances for the simulated non-engulfment pairs.

We then ran both samples through our engulfment model machinery to determine $\Delta$ln($Z$) for each simulated system. The $\Delta$ln($Z$) probability density distributions for the synthetic engulfment and non-engulfment samples are shown in the left panel of Figure \ref{fig:figure3}. 
The synthetic engulfment and non-engulfment distributions exhibit significant overlap, with $\sim$55\% of engulfment systems overlapping with the non-engulfment distribution. We conclude that our spectroscopic measurements and $\Delta$ln($Z$) analysis cannot identify nominal engulfment events (10 $M_{\oplus}$) with great confidence. We also constructed another synthetic engulfment sample drawn from our ten twin systems excluding HAT-P-4, but with 0.1$-$100 $M_{\oplus}$ added rather than 10 $M_{\oplus}$ (Figure \ref{fig:figure3}, right panel). This illustrates the $\Delta$ln($Z$) range resulting from a large set of different engulfed masses. Many of the simulated systems with $\leq$10 $M_{\oplus}$ engulfment reside to the left of the maximum $\Delta$ln($Z$) value for simulated non-engulfment systems, marked by the dashed line ($\Delta$ln($Z$) = 9.15). This further underscores that many signatures resulting from nominal 10 $M_{\oplus}$ engulfment events will not be identifiable with our $\Delta$ln($Z$) analysis. The scatter in simulated engulfed mass versus $\Delta$ln($Z$) is due to the varying stellar types of our twin systems, which result in different convective zone volumes and refractory enrichment levels for each engulfed mass amount.

\begin{figure*} 
\centering
    \vspace*{0.02in}
    \includegraphics[width=0.98\textwidth]{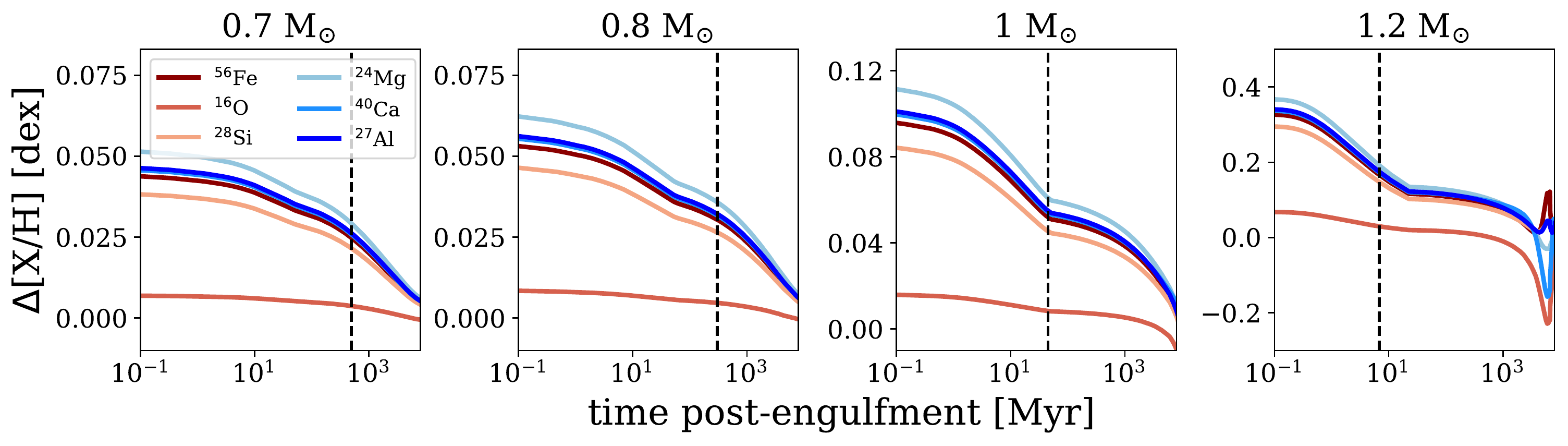}
    \caption{Evolution of the six most common isotopes in bulk Earth composition over time following engulfment of a 10 $M_{\oplus}$ bulk Earth composition planet by a 0.7$-$1.2 $M_{\odot}$ host star, as represented by abundances from \texttt{MESA} modeling. The abundances of a comparison model that did not undergo engulfment were subtracted off. The points at which the enrichments decrease to half their initial values post-engulfment range from $\sim$6$-$500 Myr depending on the engulfing star mass. These half-life points are marked by the dashed vertical black lines.}
\label{fig:figure4}
\end{figure*}

\section{Engulfment Signature Timescales} \label{sec:timescales}
Stellar interior mixing processes deplete refractory enrichments in convective zones and weaken engulfment signatures over time. The most efficient of these processes is thermohaline mixing, a form of double-diffusive convection that operates in the presence of an inverse mean-molecular-weight ($\mu$) gradient (e.g., \citealt{ulrich1972,kippenhahn1980}). Accreted planetary material is initially contained within the engulfing star's convective zone, and will create an inverse $\mu$-gradient at the convective zone base by virtue of being relatively heavy. This allows thermohaline mixing to drag engulfed material across the boundary between the convective zone and radiative stellar interior, thus attenuating photosphere refractory enrichments that compose engulfment signatures.


We ran tests with the stellar evolution code \texttt{MESA} to constrain the timescales of observable engulfment signatures considering interior mixing processes such as thermohaline instabilities. The tests involved modeling stars with masses ranging from 0.7 to 1.2 $M_{\odot}$ up to the zero-age main sequence (ZAMS), simulating engulfment of 1, 10, or 50 $M_{\oplus}$ planets via rapid accretion of bulk Earth composition material \citep{mcdonough2003}, and evolving the stars up to the end of their main sequence lifetimes. For each engulfment model, we ran another model of the same stellar mass but lacking bulk Earth accretion. The differential abundances produced by \texttt{MESA} between the engulfment and non-engulfment models thus mimic those of our binary observations. Relevant mixing processes were applied throughout these MESA runs, namely convective overshoot, elemental diffusion, radiative levitation (though we do not expect it to matter at these low stellar masses, e.g., \citealt{deal2020}), and thermohaline mixing. Thermohaline was included in these models according to the prescription of \citet{brown2013}, which provides a more accurate estimate of mixing efficiency compared to previous implementations (e.g., \citealt{kippenhahn1980}). For more details on our \texttt{MESA} modeling procedure, see Behmard et al. (\emph{in review.}) and \citet{sevilla2022}. 

We note that 10 $M_{\oplus}$ engulfment amounts can be considered nominal as runaway gas accretion is triggered by formation of a solid 10 $M_{\oplus}$ core according to the core accretion model of planet formation \citep{wuchterl2000}. Thus, most planets are expected to contain $\lesssim$10 $M_{\oplus}$ of refractory material. For 0.7 $M_{\odot}$ stars, engulfment of a 10 $M_{\oplus}$ planet does not produce observable enrichment ($\Delta$[X/H] $>$ 0.05 dex) considering the chemical dispersion observed in coeval stellar populations (0.03$-$0.05 dex, e.g., \citealt{de_silva2007,bovy2016,ness2018}). This is due to their deep convective envelopes, which heavily dilute accreted refractory material. This effect is less pronounced for more massive stars with thinner convective envelopes; solar-like stars (0.8$-$1.2 $M_{\odot}$) exhibit enrichments of $\sim$0.06$-$0.33 dex following engulfment of a 10 $M_{\oplus}$ planet. Stars in the 0.8$-$0.9 $M_{\odot}$ mass range still have moderately deep convective zones, so the initial enrichment is not significantly greater that 0.05 dex, and drops below this level after $\sim$20 Myr have passed. 1 $M_{\odot}$ stars maintain $>$0.05 dex enrichment for a longer period of $\sim$90 Myr. This timescale is still quite small compared to typical main sequence lifetimes, implying that it will be nearly impossible to detect engulfment in 1 $M_{\odot}$ stars even if it happened. Higher mass stars of 1.1$-$1.2 $M_{\odot}$ exhibit the largest and longest-lived signatures, which remain above 0.05 dex levels for $\sim$2 Gyr. Thus, these stars are the best candidates for engulfment detections. The 1.2 $M_{\odot}$ model exhibits a spike in iron photospheric abundance back to observable levels $\sim$5 Gyr after engulfment due to radiative levitation. This spike lasts for $\sim$2 Gyr, so it is possible that engulfment could also be detected in $1.2 \, M_\odot$ stars if they are observed within the window of $\sim$5$-$7 Gyr post-engulfment. However radiative levitation may be quite sensitive to stellar metallicity and poorly understood mixing processes not included in \texttt{MESA} (e.g., turbulence and rotational mixing). Thus, the $\sim$5$-$7 Gyr post-engulfment detection window for 1.2 $M_{\odot}$ stars may not be reliable. Refractory depletion behavior for 10 $M_{\oplus}$ engulfment across the 0.7$-$1.2 $M_{\odot}$ stellar mass regime is illustrated in Figure \ref{fig:figure4}. 


For cases of 1 and 50 $M_{\oplus}$ engulfment, refractory depletion patterns across different stellar masses are similar to those of 10 $M_{\oplus}$ engulfment, but scaled down and up, respectively. For 1 $M_{\oplus}$ engulfment, stars with masses in the range 0.7$-$1.1 $M_{\odot}$ begin with enrichment levels at $\lesssim$0.05 dex, and thus never exhibit detectable engulfment signatures. However 1.2 $M_{\oplus}$ stars begin with $>$0.05 dex enrichment, and maintain this level for $\sim$100 Myr. For 50 $M_{\oplus}$ engulfment, 0.7$-$1.2 $M_{\odot}$ stars maintain $>$0.05 dex enrichment for $\sim$3$-$8 Gyr. However, planets containing up to 50 $M_{\oplus}$ of refractory material are predicted to be quite rare \citep{batygin2016}.

We ran additional \texttt{MESA} models with different engulfing star and accretion conditions, and found that observable signature timescales increase for sub-solar metallicities, or if engulfment occurs at later times post-ZAMS. Engulfment of a 10 $M_{\oplus}$ planet by a 1 $M_{\odot}$ sub-solar ($Z$ = 0.012) metallicity star results in $>$0.05 dex refractory enrichment for $\sim$3 Gyr. This is due to two effects: refractory enrichments are highlighted in low metallicity environments, and stars with low metallicities have thinner convective envelopes. Engulfment events occurring 300 Myr$-$3 Gyr post-ZAMS also yield signatures that remain observable on $>$1 Gyr timescales; 10 $M_{\oplus}$ engulfment by a 1 $M_{\odot}$ star at these times produces $>$0.05 dex enrichment that lasts for $\sim$1.5 Gyr. Such late-stage engulfment results in longer observable signature timescales because refractory depletion via thermohaline is suppressed due to a counteracting positive $\mu$-gradient from helium settling over time. Still, these timescales are short compared to main sequence lifetimes; our \texttt{MESA} results imply that enrichment from nominal 10 $M_{\oplus}$ engulfment events will rarely be observable in solar-like stars that are several Gyr old.

\subsection{Twin Importance} \label{sec:twin_importance}
As mentioned in Section \ref{sec:sample}, binary twin systems are well suited for engulfment surveys because twin companions are at the same evolutionary stage. Our \texttt{MESA} results underscore this; stars with different masses and evolutionary states exhibit different rates of refractory depletion, and \citet{sevilla2022} found this to be true even in the absence of engulfment due to diffusion (\citealt{sevilla2022}, Figure 9). This implies that non-twin binary pair stars will always have different refractory abundances, with differences increasing in time. Thus, only twin systems are capable of yielding reliable planet engulfment signatures. For a full description of our \texttt{MESA} modeling analysis and results, see Behmard et al. (\emph{in review}).


\begin{deluxetable*}{lrrrrrr}
\centering
\tablewidth{0.7\textwidth}
\tabletypesize{\footnotesize}
\tablewidth{0pt}
\tablecaption{Engulfment Model Parameters \label{tab:table7}}
\tablecolumns{7}
\tablehead{
\colhead{Binary System} &
\colhead{sep} &
\colhead{$M$} &
\colhead{$\sigma_{\textrm{jit}}$} &
\colhead{shift} &
\colhead{flat model shift} &
\colhead{$\Delta$ln$(Z)$} \\
\colhead{} &
\colhead{AU} &
\colhead{$M_{\oplus}$} &
\colhead{} &
\colhead{dex} &
\colhead{dex} &
\colhead{} 
}
\startdata
HD 99491-92 &    510 $\pm$ 0.30 &  11.73 $\pm$  2.98 &   0.02 $\pm$ 0.01 &  $-$0.08 $\pm$ 0.01  & $-$0.04 $\pm$ 0.01 &   5.21 \\
Kepler-477* &    560 $\pm$ 5.9 &   3.06 $\pm$  0.85 &   0.03 $\pm$ 0.01 &  $-$0.10 $\pm$ 0.02  & $-$0.05 $\pm$ 0.01 &    4.53 \\
Kepler-515* &    650 $\pm$ 2.2 &   8.62 $\pm$  2.92 &   0.08 $\pm$ 0.01 &  $-$0.08 $\pm$ 0.03  & $-$0.02 $\pm$ 0.02 &    3.98 \\
WASP-180 &   1200 $\pm$ 6.2 &   4.93 $\pm$  2.23 &   0.09 $\pm$ 0.02 &  $-$0.07 $\pm$ 0.02  & $-$0.03 $\pm$ 0.02 &    2.31 \\
WASP-94*$^{\dagger}$ &   3200 $\pm$ 17 &   2.95 $\pm$  1.55 &   0.03 $\pm$ 0.01 &  $-$0.03 $\pm$ 0.01  & 0.00 $\pm$ 0.01 &    1.96 \\
HAT-P-4*$^{\dagger}$ &  30000 $\pm$ 140 &   5.60 $\pm$  1.64 &   0.03 $\pm$ 0.01 &   0.05 $\pm$ 0.01  &  0.10 $\pm$ 0.01 &   1.82 \\
Kepler-25 &   2000 $\pm$ 5.5 &  8.36 $\pm$  5.04 &   0.10 $\pm$ 0.02 &  $-$0.03 $\pm$ 0.03  &  0.01 $\pm$ 0.02 &    1.46 \\
HD 133131*$^{\dagger}$ & 380 $\pm$ 0.62 &   0.50 $\pm$  0.23 &   0.00 $\pm$ 0.00 &  0.00 $\pm$ 0.01  &  0.01 $\pm$ 0.00 &  1.04 \\
WASP-160* &   8300 $\pm$ 28 &   4.36 $\pm$  2.70 &   0.01 $\pm$ 0.01 &  0.02 $\pm$ 0.01  & 0.04 $\pm$ 0.01 &   0.98 \\
WASP-64$^{\dagger}$* &   8700 $\pm$ 34 &   1.83 $\pm$  0.89 &   0.00 $\pm$ 0.00 &  0.00 $\pm$ 0.01  &  0.01 $\pm$ 0.00 &   0.66 \\
HD 106515*$^{\dagger}$ &    230 $\pm$ 0.25 &   0.97 $\pm$  0.84 &   0.00 $\pm$ 0.00 &  $-$0.02 $\pm$ 0.01  & $-$0.01 $\pm$ 0.00 &   0.01 \\
WASP-173 &   1400 $\pm$ 6.9 &   6.18 $\pm$  3.20 &   0.02 $\pm$ 0.01 &  $-$0.10 $\pm$ 0.01  & $-$0.07 $\pm$ 0.01 &    $-$0.12 \\
K2-27* &   8100 $\pm$ 31 &   0.86 $\pm$  0.70 &   0.02 $\pm$ 0.01 &   $-$0.02 $\pm$ 0.01  &  $-$0.02 $\pm$ 0.00 &   $-$0.25 \\
HD 178911* &    650 $\pm$ 0.39 &   1.62 $\pm$  1.81 &   0.04 $\pm$ 0.01 &  0.00 $\pm$ 0.01  &  0.00 $\pm$ 0.01 &   $-$0.27 \\
HD 132563*$^{\dagger}$ &    430 $\pm$ 0.52 &   0.42 $\pm$  0.29 &   0.02 $\pm$ 0.01 &  $-$0.02 $\pm$ 0.01  & $-$0.01 $\pm$ 0.00 &   $-$0.37 \\
HD 40979 &   6500 $\pm$ 4.5 &   2.45 $\pm$  3.10 &   0.07 $\pm$ 0.01 &  $-$0.03 $\pm$ 0.02  & $-$0.02 $\pm$ 0.01 &   $-$0.40 \\
WASP-127* &   6500 $\pm$ 20 &   0.62 $\pm$  0.42 &   0.03 $\pm$ 0.01 &   0.03 $\pm$ 0.01  &  0.04 $\pm$ 0.01 &   $-$0.55 \\
Kepler-99* &   3100 $\pm$ 8.5 &   1.16 $\pm$  1.51 &   0.04 $\pm$ 0.01 &  0.00 $\pm$ 0.01  & 0.00 $\pm$ 0.01 &   $-$0.55 \\
HD 80606-07* &   1400 $\pm$ 1.6 &   0.66 $\pm$  0.71 &   0.01 $\pm$ 0.00 &   $-$0.01 $\pm$ 0.00  &  $-$0.01 $\pm$ 0.00 &   $-$0.64 \\
KELT-2 &    320 $\pm$ 0.83 &  1.33 $\pm$  1.36 &   0.20 $\pm$ 0.03 &  0.00 $\pm$ 0.03  & 0.02 $\pm$ 0.04 &   $-$0.75 \\
HAT-P-1*$^{\dagger}$ &   1800 $\pm$ 4.1 &   0.65 $\pm$  0.61 &   0.03 $\pm$ 0.01 &   0.02 $\pm$ 0.01  &  0.03 $\pm$ 0.01 &   $-$0.76 \\
XO-2*$^{\dagger}$ &   4700 $\pm$ 11 &   9.46 $\pm$  4.97 &   0.04 $\pm$ 0.01 &  0.04 $\pm$ 0.01  & 0.06 $\pm$ 0.01 &   $-$0.83 \\
Kepler-104* &   6900 $\pm$ 27 &   0.68 $\pm$  0.45 &   0.13 $\pm$ 0.02 &  $-$0.06 $\pm$ 0.02  & $-$0.07 $\pm$ 0.02 &   $-$1.44 \\
16 Cyg*$^{\dagger}$ &    840 $\pm$ 0.28 &   0.29 $\pm$  0.22 &   0.00 $\pm$ 0.00 & 0.02 $\pm$ 0.00  & 0.03 $\pm$ 0.00 &   $-$1.50 \\
HD 20781-82* &   9100 $\pm$ 7.9 &   0.25 $\pm$  0.23 &   0.01 $\pm$ 0.00 &  0.01 $\pm$ 0.00  & 0.01 $\pm$ 0.00 &   $-$1.70 \\
Kepler-1063  &    580 $\pm$ 27 &   1.50 $\pm$  1.12 &   0.00 $\pm$ 0.00 &   0.06 $\pm$ 0.01  &  0.07 $\pm$ 0.00 &   $-$1.79 \\
KELT-4 &    340 $\pm$ 1.2 &  3.55 $\pm$  4.74 &   0.31 $\pm$ 0.04 &   $-$0.03 $\pm$ 0.03  &  $-$0.09 $\pm$ 0.06 &   $-$2.05 \\
HD 202772*$^{\dagger}$ &    210 $\pm$ 1.7 &   1.13 $\pm$  1.05 &   0.14 $\pm$ 0.02 &   0.03 $\pm$ 0.03  &  0.08 $\pm$ 0.03 &   $-$2.20  \\
$\omega$ Ser* &   5700 $\pm$ 35 &  48.09 $\pm$ 18.58 &   0.20 $\pm$ 0.03 &   0.03 $\pm$ 0.03  &  0.20 $\pm$ 0.04 &   $-$4.07
\enddata
\tablecomments{This table lists the binary separation, modeled amount of engulfed planetary mass, fitted jitter term $\sigma_{\textrm{jit}}$, engulfment model shift, flat model shift, and difference in engulfment model and flat model Bayesian evidence $\Delta$ln$(Z)$ for each of the remaining 29 binary pairs in the engulfment sample. For each pair, we chose either the planet host or the non-planet host to be the engulfing star based on which order yielded the largest $\Delta$ln$(Z)$. Pairs where the planet host was assumed to be the engulfing star are marked with *, and stellar twin systems ($\Delta$$T_{\textrm{eff}}$ $<$ 200 K) are marked with $\dagger$. The binary pairs are sorted by $\Delta$ln$(Z)$.}
\end{deluxetable*}

\section{Engulfment or Primordial Differences} \label{sec:potential_detections}

Before presenting our results, we outline our criteria for engulfment:

\begin{enumerate}

    \item The stellar companions qualify as twins ($\Delta$$T_{\textrm{eff}}$ $<$ 200 K, \citealt{andrews2019}).
    
    \item There is a large ($\geq$10 $M_{\oplus}$) amount of recovered engulfed mass from our model, with larger mass amounts considered more robust (Section \ref{sec:lnZ_criteria}).
    
    \item The engulfment model shift (base of the $T_c$ pattern across all abundances) lies above $-$0.05 dex. This is justified because the amount of primordial chemical dispersion between bound stellar companions is not expected to exceed 0.03$-$0.05 dex (e.g., \citealt{de_silva2007,bovy2016,ness2018}), and engulfment will result in a positive addition to the differential abundances.
    
    \item These previous two conditions are satisfied across removal of each abundance, tested via applying the engulfment model after removing one abundance at a time. This leave-one-out test ensures that the $T_c$ trends are not driven by any single abundance.
    
    \item There is a positive $\Delta$A(Li) between stellar companions, in the direction of potential engulfment.
    
    
\end{enumerate}

\begin{figure*} 
\centering
    \includegraphics[width=0.98\textwidth]{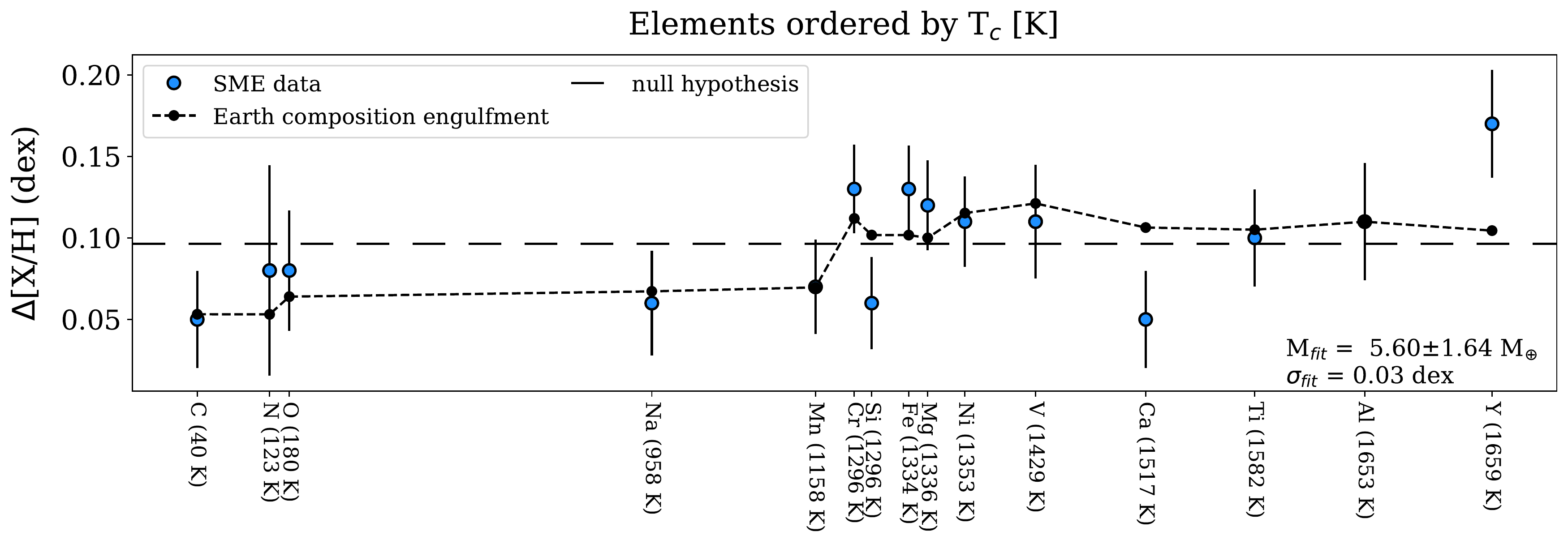}
\caption{Fitted model to the differential abundance measurements between the hot Jupiter host HAT-P-4 and its companion. As in Fig. \ref{fig:figure2}, the blue circles represent the abundance differences, and the black dots are our model fit with 5.60 $\pm$ 1.64 $M_\oplus$ of bulk Earth composition engulfed material added to the convective zone of one star. The amount of modeled engulfed material and fitted scatter is provided in the lower right corner of the plot. The null hypothesis (no engulfment, but uniform abundance enrichment across all elements for one star) is shown by the long-dash line.}
\label{fig:figure5}
\end{figure*}

In light of our \texttt{MESA} results, we only considered the eleven twin systems in our sample as potential engulfment detections. However, we still applied our engulfment model to all 29 systems. Because we did not know which star in each pair may have undergone engulfment, both cases were considered for each system. All $\Delta$ln($Z$) measurements for our engulfment sample are reported in Table \ref{tab:table7}. Among our eleven twin systems, only HAT-P-4 exhibits a positive Bayesian evidence difference ($\Delta$ln($Z$) = 1.82) and an engulfment model shift that lies above $-$0.05 dex (Figure \ref{fig:figure5}). The amount of recovered mass is 5.60 $\pm$ 1.64 $M_\oplus$, and remains above 5.11 $\pm$ 1.72 $M_\oplus$ across removal of each abundance. The HAT-P-4 $\Delta$ln($Z$) value of 1.82 is well below our suggested cutoff of $\Delta$ln($Z$) = 9.15 justified by our Bayesian evidence analysis (Section \ref{sec:lnZ_criteria}). Still, HAT-P-4 satisfies more of our engulfment claim criteria than any other system in our sample, making it the most promising potential engulfment detection. We note that there are five other systems (HD 99491-92, Kepler-477, Kepler-515, WASP-180, and WASP-94) with $\Delta$ln($Z$) above the HAT-P-4 value of $\sim$1.82 (Table \ref{tab:table7}), but none satisfy the model shift above $-$0.05 dex criterion, and four do not qualify as twins (HD 99491-92, Kepler-477, Kepler-515, and WASP-180).

\begin{figure}[t]
    \centering
        \includegraphics[width=0.5\textwidth]{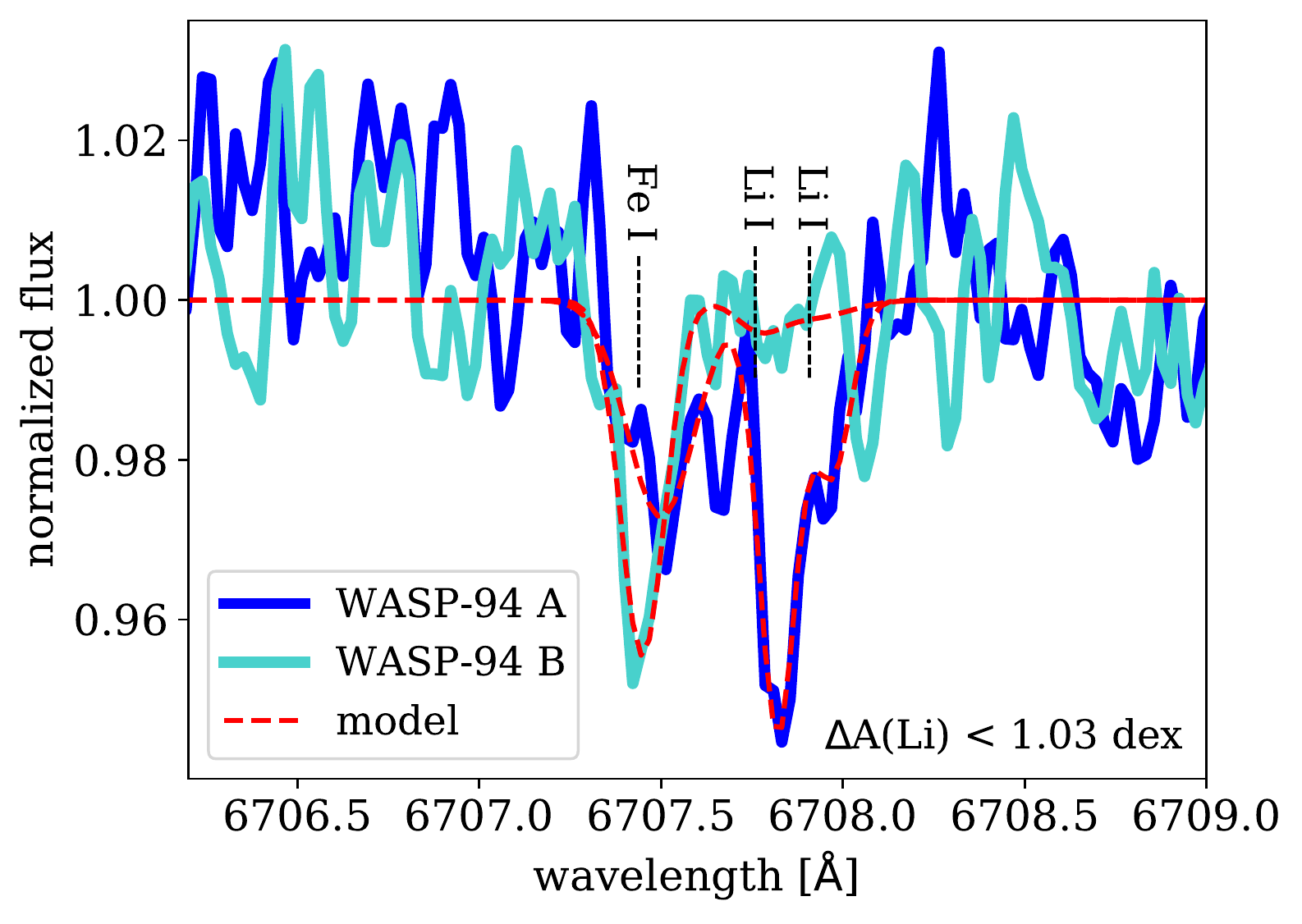} 
    \caption{Li doublet region for WASP-94, with the normalized spectra of the stellar companions with lower and higher Li abundance plotted in light and dark blue, respectively. The model fits used to derive Li EWs and abundances are illustrated by the red dashed lines. The Fe I and Li I transitions are marked, and the differential Li abundance is provided in the lower right corner.}
\label{fig:figure6}
\end{figure}

\begin{figure*}[t]
    \centering
    \begin{minipage}{0.5\textwidth}
        \centering
        \includegraphics[width=0.99\textwidth]{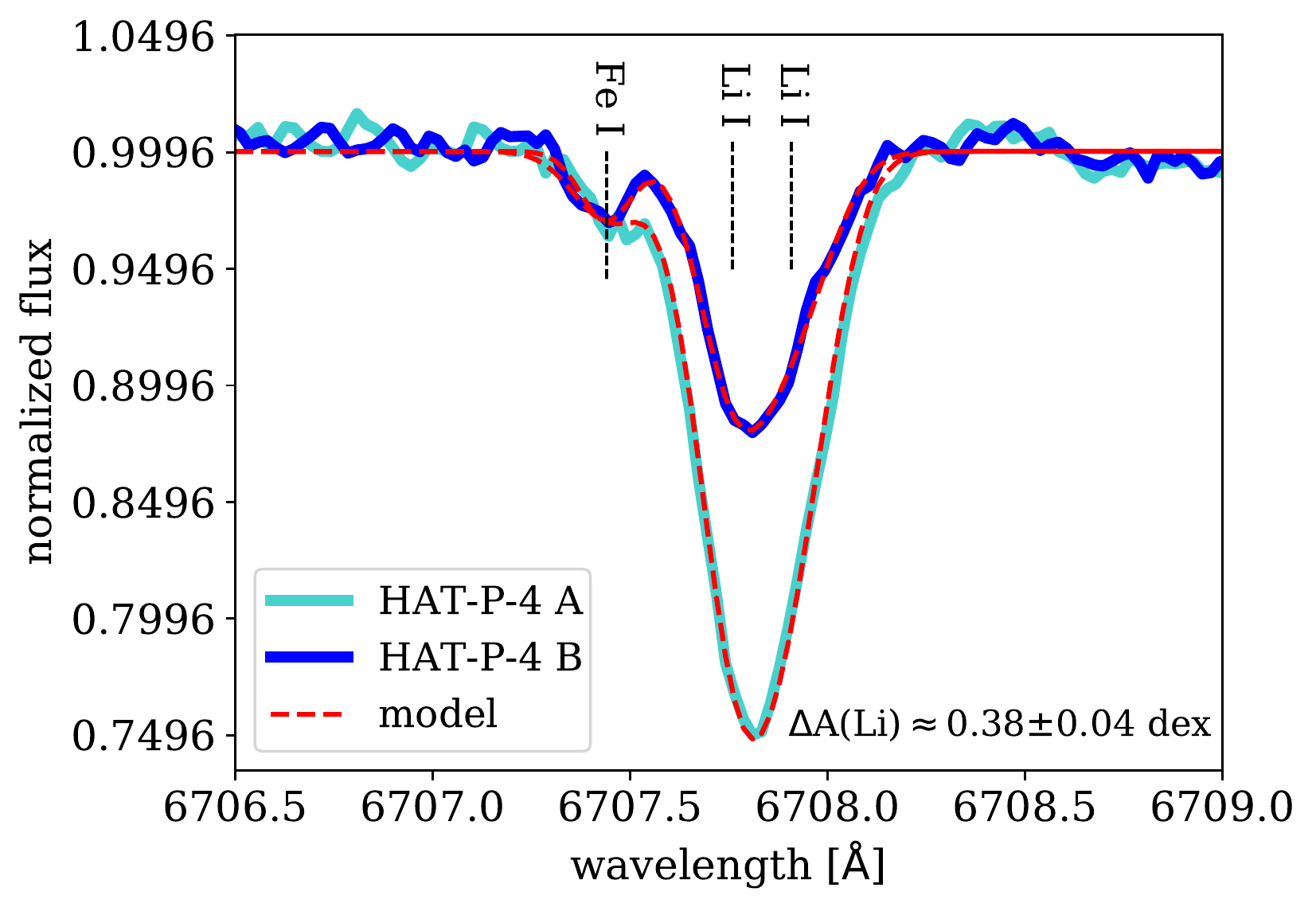} 
    \end{minipage}\hfill
    \begin{minipage}{0.465\textwidth}
        \centering
        \includegraphics[width=0.999\textwidth]{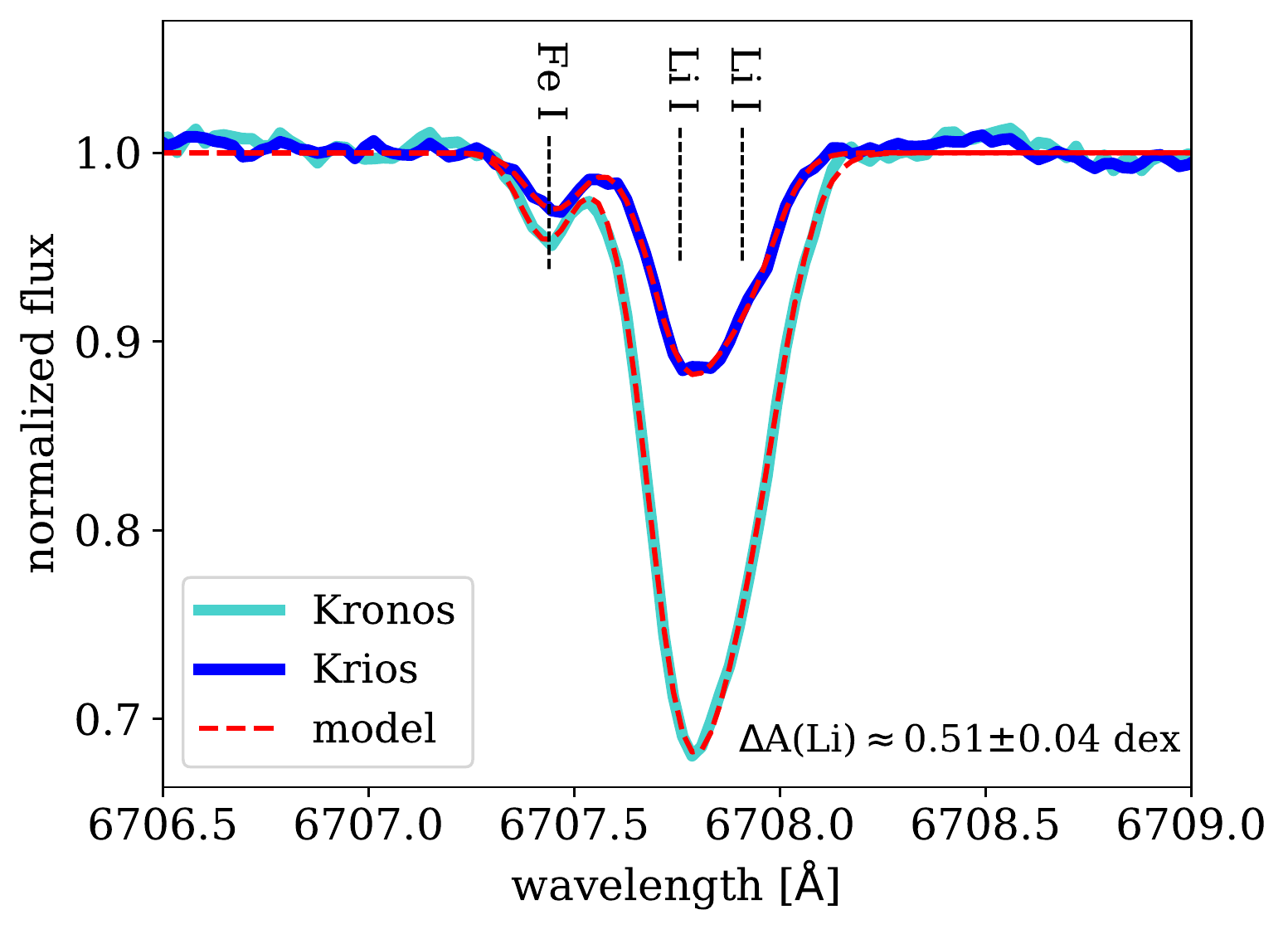} 
    \end{minipage}
    \caption{Li doublet regions for HAT-P-4 (left) and Kronos-Krios (right), with the normalized spectra of the stellar companions with lower and higher Li abundance plotted in light and dark blue, respectively. The model fits used to derive Li EWs and abundances are illustrated by the red dashed lines. The Fe I and Li I transitions are marked, and the differential Li abundances are provided in the lower right corners of the panels.}
\label{fig:figure7}
\end{figure*}

There are five systems in our sample with $\Delta$Li $>$ 0.1 dex and $\Delta$ln($Z$) $>$ 1.82, of which only two (HAT-P-4 and WASP-94) are twin binaries. HAT-P-4 and WASP-94 have Li abundances differences between the stellar companions of $\Delta$A(Li) $\approx$ 0.38 $\pm$ 0.04 dex and $\Delta$A(Li) $<$1.03 dex, respectively (we only report the upper limit $\Delta$A(Li) value for WASP-94 because the Li EW is smaller than its associated error for WASP-94 B). The WASP-94 Li doublet appears quite weak (Figure \ref{fig:figure6}). Thus, we argue that only HAT-P-4 has a $\Delta$A(Li) potentially indicating engulfment. Kronos-Krios has a Li abundance difference of $\Delta$A(Li) $\approx$ 0.51 $\pm$ 0.04 dex, which is comparable to the Li abundance difference of HAT-P-4. We plot the Li doublet regions for these systems in Figure \ref{fig:figure7}.

To claim engulfment, we also need to verify that the differential abundance pattern supporting an engulfment scenario is not the result of primordial chemical differences between the two stellar companions. This was investigated via binary companion separations; chemical gradients could potentially increase with distance in molecular clouds, resulting in varied chemistry between widely separated stellar siblings. Thus, we must consider the possibility that large differential abundances in wide binary systems may result from primordial chemical differences rather than planet engulfment. There is some observational evidence for this possibility from open clusters, whose stars are widely separated by definition. \citet{ness2018} examined pairs of red giants in seven open clusters, and found that a minority of pairs are highly chemically dissimilar according to a measure of chemical distance between the companions for 20 elements of $\chi^{2}$ $\approx$ 70. For reference, most of the intra-cluster pairs are chemically homogeneous and exhibit $\chi^{2}$ $\approx$ 20, corresponding to typical abundance dispersions of $\sim$0.03 dex. \citet{liu2016b} put forward possibilities to explain such abundance differences in open clusters, such as supernova ejection in the proto-cluster cloud, or pollution of metal-poor gas. Both are contingent upon insufficient turbulent mixing within the cloud that would fail to smooth out chemical inhomogeneities.

To examine the possibility that the abundance differences of our systems are primordial, we calculated the projected separations for our 29 planet host binaries with \texttt{SME}-determined $T_{\textrm{eff}}$ $>$ 4700 K using \emph{Gaia} Data Release 3 (DR3) astrometry. The errors on projected separations were taken as the scatter in calculated separations after sampling from the astrometric data uncertainty distributions 100 times for each system. These separations are reported in Table \ref{tab:table7}. The projected separation of HAT-P-4 is 30,000 $\pm$ 140 AU, which is larger than that of any other binary in our sample by an order of magnitude (Table \ref{tab:table7}). The projected separation can be considered a factor of $\sqrt{1.5}$ smaller than the true distance, and results in a value that exceeds typical turbulence scales in molecular clouds (0.05$-$0.2 pc, \citealt{brunt2009} and references therein). This indicates that the HAT-P-4 stellar companions may have formed in distinct areas of chemodynamical space within their birth cloud. Thus, we regard the HAT-P-4 differential abundance pattern as potentially due to primordial chemical differences between the two stars rather than planet engulfment.


\section{Assessment of Published Systems} \label{sec:previous_systems}

There are ten planet host binary systems with high-precision abundances previously measured (HAT-P-1, HD 20781-82, XO-2, WASP-94, HAT-P-4, HD 80606-07, 16-Cygni, HD 133131, HD 106515, WASP-160; Table \ref{tab:table1}). Depending on the study, four to six of these systems are claimed as engulfment detections. Because no potential engulfment signatures were found in our sample aside from HAT-P-4, we were interested in testing if previously reported datasets for these ten systems yield robust signatures according to our engulfment model. 

We found that six of the systems exhibit $\Delta$ln($Z$) $>$ 1.82, above HAT-P-4 (16 Cygni, XO-2, HD 20781-82, HD 133131, WASP-94, and WASP-160). However this depends on the reported dataset; the 16 Cygni abundances derived by \citet{tucci_maia2014}, \citet{tucci_maia2019}, and \citet{ryabchikova2022} are above this cutoff, but those of \citet{ramirez2011} yield a negative $\Delta$ln($Z$). Likewise, the XO-2 abundances derived by \citet{ramirez2015} and \citet{biazzo2015} pass the HAT-P-4 cutoff, but those of \citet{teske2015} yield a negative $\Delta$ln($Z$). The \citet{ramirez2011} and \citet{teske2015} studies did not claim engulfment. Our fitted engulfment model to the \citet{ryabchikova2022} 16 Cygni dataset also exhibits a shift below $-$0.05 dex, which violates our engulfment criteria. This is also true for the \citet{mack2014} and \citet{teske2016} datasets for HD 20781-82 and WASP-94, respectively. The \citet{teske2016b} HD 133131 dataset passes this engulfment model shift criterion, but yields a small engulfed mass estimate ($M$ = 1.13 $\pm$ 0.51 $M_{\oplus}$), and is not claimed as engulfment by \citet{teske2016b}. This leaves the \citet{jofre2021} WASP-160 dataset, which yields an estimated engulfed mass of $M$ = 7.73 $\pm$ 1.59 $M_{\oplus}$ and $\Delta$ln($Z$) = 9.37. However WASP-160 is part of our sample, and our \texttt{SME} abundances do not clearly favor an engulfment scenario ($\Delta$ln($Z$) = 0.98). We conclude that there is no evidence for strong engulfment detections in the literature aside from potentially Kronos-Krios.

\begin{figure*}[t]
    \centering
    \begin{minipage}{0.49\textwidth}
    \vspace{1mm}
        \centering
        \includegraphics[width=0.999\textwidth]{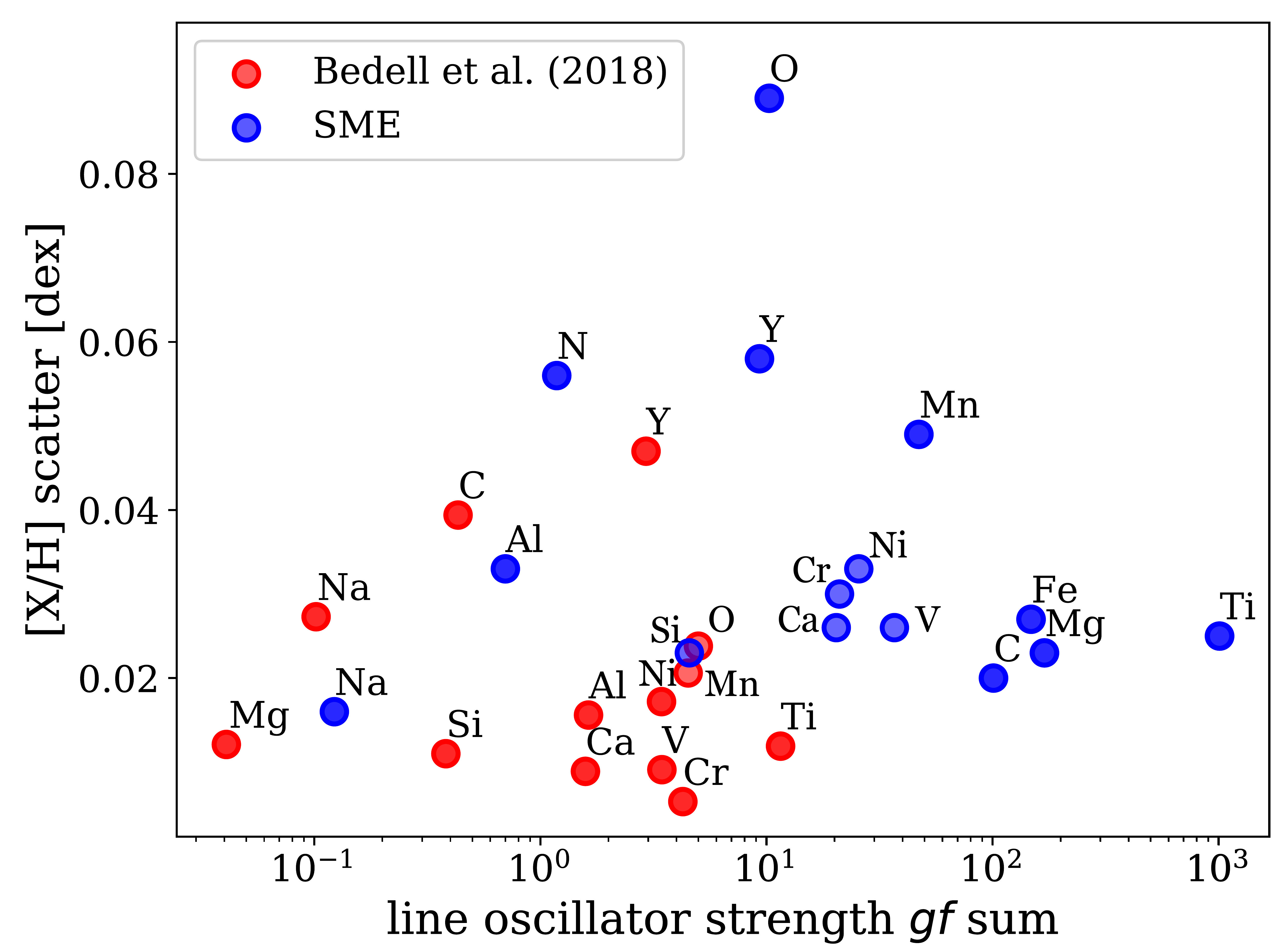} 
    \end{minipage}\hfill
    \begin{minipage}{0.475\textwidth}
        \centering
        \includegraphics[width=0.999\textwidth]{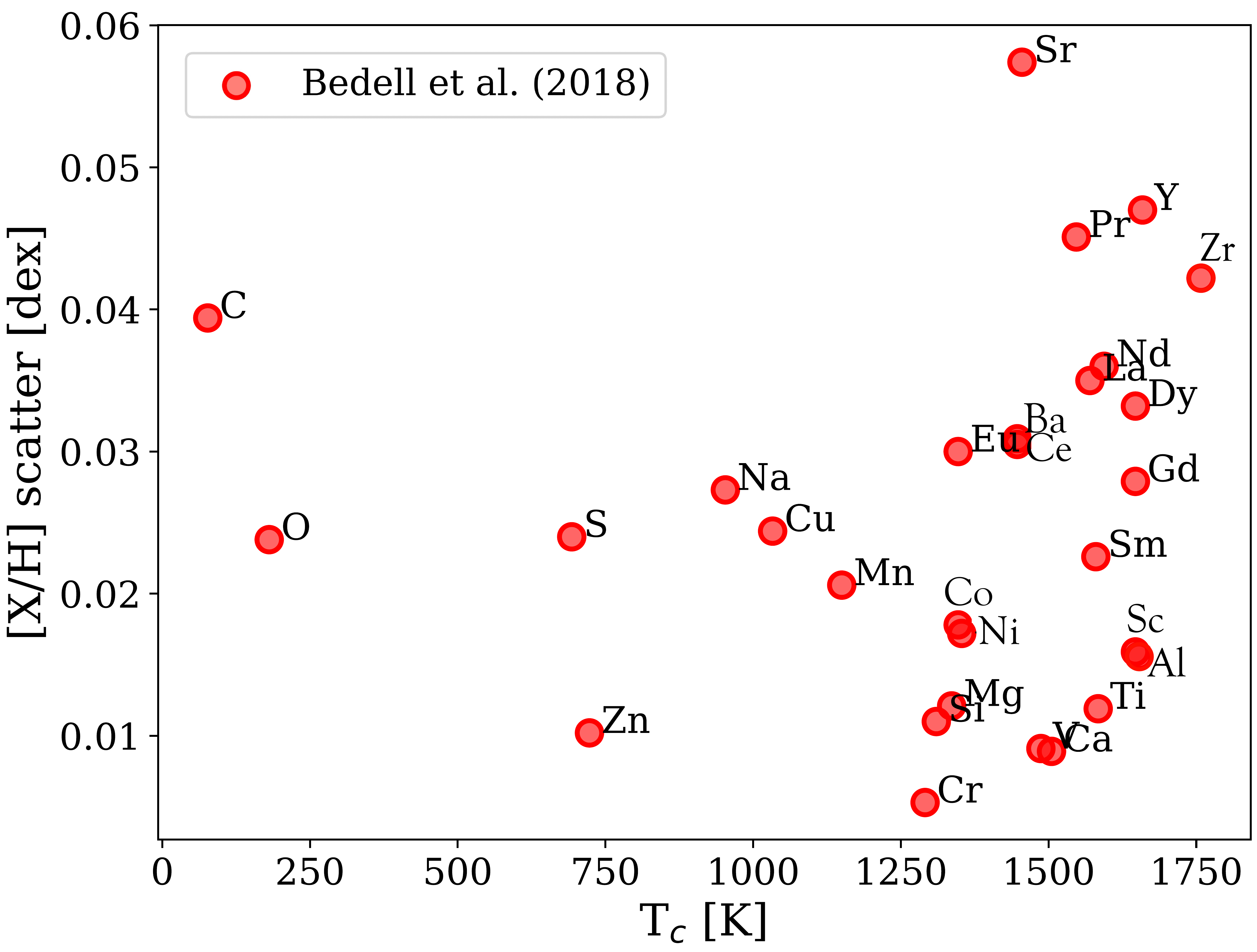} 
    \end{minipage}
    \caption{The left panel displays the scatter in abundance measurements vs. the sum of oscillator strengths $gf$ for each element in the \texttt{SME} line list and our engulfment sample (blue points), and for the \citet{bedell2018} line list and solar twin sample (red points). The abundance scatter between companions in our binary sample and across the sample of solar twins increases as the included lines for each abundance become fewer and weaker. The right panel displays the scatter in abundance measurements vs. $T_c$ for the \citet{bedell2018} sample considering all elements reported in their study.}
\label{fig:figure8}
\end{figure*}

\subsection{Abundance Scatter} \label{sec:abund_scatter}
Abundance discrepancies between different studies of the same stars can be attributed to usage of different instruments (e.g., \citealt{bedell2014}); differences in the acquired spectra such as varying SNR levels (e.g., \citealt{liu2018}); or to differences in abundance measurement pipelines that may employ different spectral synthesis codes, continuum placement, EW measurement procedures, and line lists (e.g., \citealt{schuler2011,liu2018}). A few studies that exemplify these discrepancy sources are \citet{saffe2015}, \citet{mack2016}, and \citet{liu2018}, which all analyzed HD 80606-07, but derived widely varying abundance measurements. \citet{saffe2015} and \citet{mack2016} used the same set of Keck-HIRES observations, but derived abundances that often do not agree within their combined uncertainties at the 1$\sigma$ level. \citet{liu2018} obtained higher SNR observations of HD 80606-07, and claimed that their abundance measurements are more reliable because their average uncertainties ($\sim$0.007 dex) are much smaller than those of \citet{saffe2015} and \citet{mack2016} (0.02 dex and 0.027 dex, respectively).

These three studies also employed different line lists. The \citet{saffe2015} list includes the highest number of lines at $\sim$500, followed by the \citet{liu2018} list with $\sim$250 lines, then the \citet{mack2016} list with $\sim$125 lines. To quantify the quality of these different line lists, we calculated the summed oscillator strength $gf$ over each line corresponding to a single abundance. As expected, this quantity is a factor of 2$-$4 higher for the \citet{liu2018} and \citet{saffe2015} line lists compared to the \citet{mack2016} line list averaging across all abundances. This is likely responsible for the approximate abundance measurement agreement between \citet{liu2018} and \citet{saffe2015}, but not \citet{mack2016}. For comparison, the line list we employed in our \texttt{SME} analysis includes over 7500 lines, making the summed $gf$ quantity $\sim$100 times higher than that of the \citet{saffe2015} line list. The average difference for our \texttt{SME}-derived HD 80606-07 abundances is +0.006, also in better agreement with \citet{liu2018} and \citet{saffe2015} compared to \citet{mack2016}.




We were interested in quantifying how line lists affect abundance measurements by examining if abundance prediction scatter changes as a function of line number and strength. We tested our \texttt{SME} line list against the abundance scatter between companions in the ten twin systems excluding HAT-P-4 from our engulfment sample, and found that abundances with fewer and weaker lines according to oscillator strength $gf$ (e.g., O, Y, N) exhibit larger abundance prediction scatter (Figure \ref{fig:figure8}, left panel, blue points). This indicates that scatter is large for volatile and highly refractory abundances that anchor the lower and upper portions of the $T_c$ trend, respectively. We carried out the same analysis for the \citet{bedell2018} sample of solar twins and the line list used in their \texttt{MOOG} analysis, and found the same trend of abundance scatter increasing with fewer and weaker lines per abundance (Figure \ref{fig:figure8}, left panel, red points). We also examined the \citet{bedell2018} abundance scatter as a function of $T_c$, and found that abundances with low (e.g., C and O) and high $T_c$ (e.g., Zr and Y) exhibit large scatter similar to our \texttt{SME} results (Figure \ref{fig:figure8}, right panel). These findings show that large line lists with strong spectral features are necessary for measuring precise abundances, and elements that anchor the $T_c$ trend lack an abundance of strong features and thus exhibit large scatter. This is unsurprising for the low $T_c$ abundances; volatile elements like C, N, and O are often locked in molecular species that create blended features, making it difficult to identify strong, well-isolated lines. Because elements important for establishing a $T_c$ trend tend to have large uncertainties, we expect that a $T_c$ pattern can occur randomly in the absence of engulfment.

\section{Discussion} \label{sec:discussion}
\begin{figure}[t]
    \centering
        \includegraphics[width=0.475\textwidth]{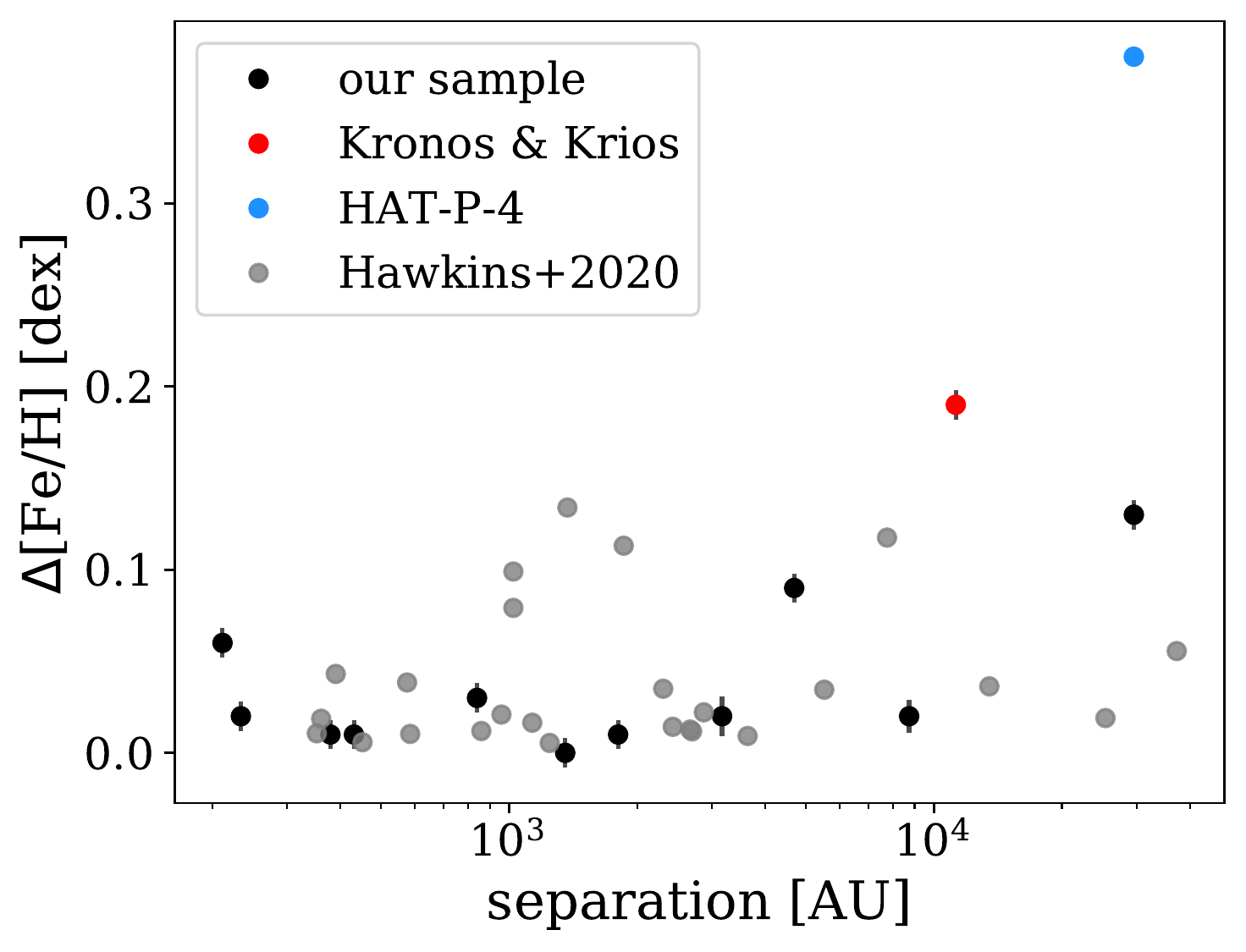} 
    \caption{$\Delta$[Fe/H] (right) vs. projected binary separation of the eleven twin systems in our sample assessed for engulfment signatures (black). The 25 wide binaries from \citet{hawkins2020} are plotted for comparison (gray), along with the points representing Kronos-Krios (red) and HAT-P-4 (blue). There appears to be a trend of increasing $\Delta$[Fe/H] as a function of separation across all samples. The \citet{spina2021} systems are not explicitly shown because most of their systems that qualify as twins are drawn from the \citet{hawkins2020} sample, and the remaining do not have reported separations.}
\label{fig:figure9}
\end{figure}

We did not recover any strong planet engulfment detections in our planet host binary sample. HAT-P-4 is the only system whose abundances exhibit a possible engulfment signature. This binary is composed of two solar-like (G0V + G2V) stars, with the primary hosting a 0.68 $M_{\textrm{Jup}}$ hot Jupiter at an orbital period of $\sim$3 days \citep{kovacs2007}. Our engulfment model recovers 5.60 $\pm$ 1.64 $M_{\oplus}$ of engulfed mass by the planet host star. However, HAT-P-4's $\Delta$ln($Z$) value of $\sim$1.82 does not strongly support an engulfment claim, and the system sustains only $\Delta$ln($Z$) = 1 across the leave-one-out abundance test. For reference, these values are well below the maximum $\Delta$ln($Z$) value of our synthetic non-engulfment systems (9.15, Section \ref{sec:lnZ_criteria}), indicating that the HAT-P-4 engulfment signature could be a false positive. In addition, the projected separation of HAT-P-4 (30,000 $\pm$ 140 AU) is an order of magnitude larger than that of any other binary in our sample (Table \ref{tab:table7}), and exceeds the lower bound of typical turbulence scales in molecular clouds (0.05$-$0.2 pc, \citealt{brunt2009} and references therein). This suggests that HAT-P-4 A and B formed far from each other within their birth cloud, and were separated by large chemical gradients that gave rise to the differential abundance pattern we see today. In this case, the chemical differences of HAT-P-4 would be primordial rather than the result of planet engulfment. It is possible that the Kronos-Krios abundance pattern is also primordial; we calculated the projected separation for this system to be 11,000 $\pm$ 12 AU. There is also a tentative trend of increasing abundance difference as a function of binary separation in our sample of eleven twin systems. To illustrate this, we plot their $\Delta$[Fe/H] as a function of separation in Figure \ref{fig:figure9}, along with those of the 25 wide binaries from \citet{hawkins2020}.

\begin{figure}[t]
    \centering
        \includegraphics[width=0.48\textwidth]{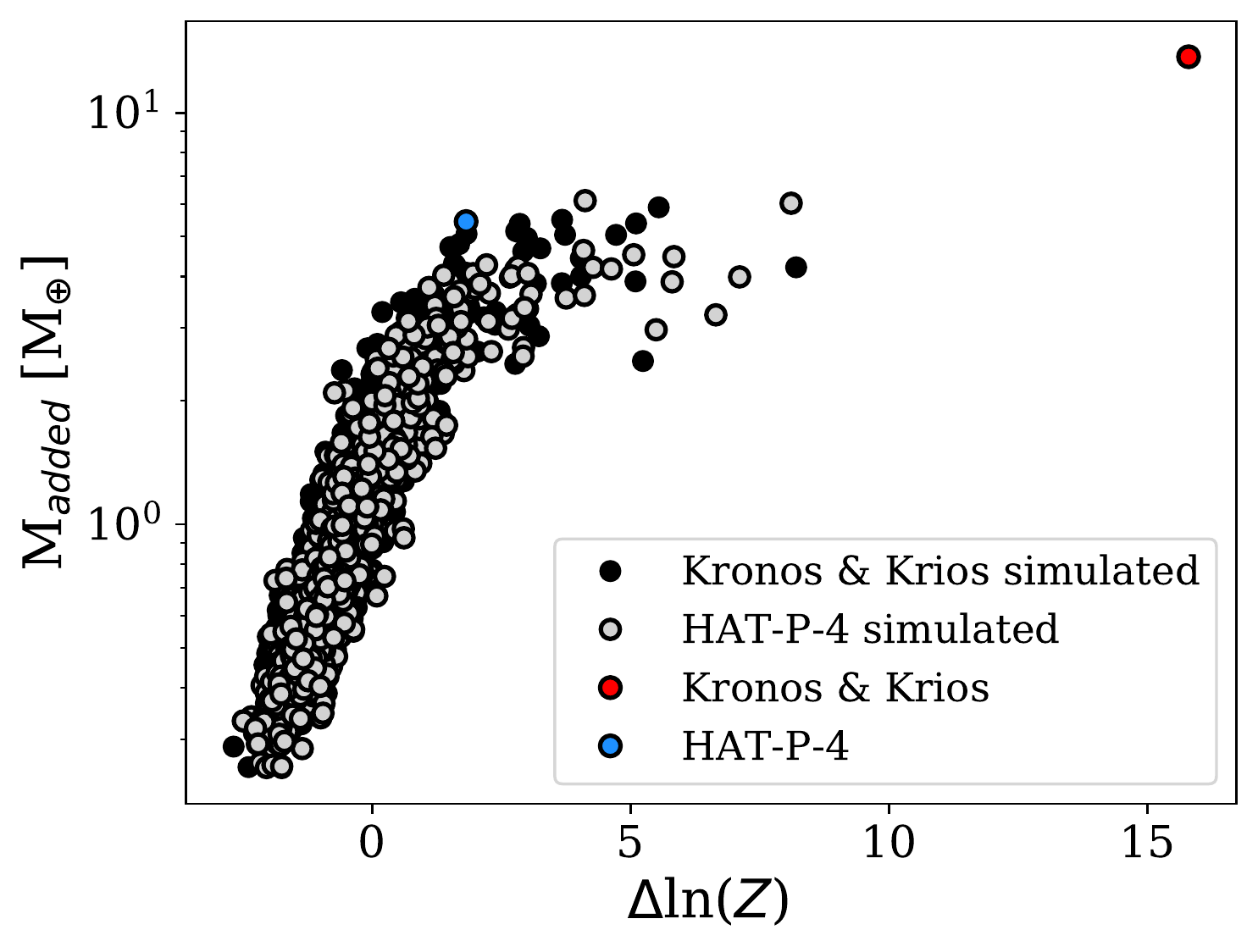} 
    \caption{Estimated amounts of engulfed material from our engulfment model fits vs. $\Delta$ln($Z$) values for the 1000 simulated systems assuming the HAT-P-4 companion masses and convective zones (gray), and the 1000 simulated systems assuming those of Kronos-Krios (black). The abundances of the simulated systems were drawn from Gaussian distributions with widths equal to the average abundance scatter per element of our ten twin systems excluding HAT-P-4. The real data for Kronos-Krios and HAT-P-4 are also shown for comparison as the red and blue dots, respectively. The simulated systems can mimic the $T_c$ trend of HAT-P-4, but not that of Kronos-Krios.}
\label{fig:figure10}
\end{figure}

While a $T_c$-dependent abundance pattern is a signpost of planet engulfment, it is possible that the $T_c$-dependent patterns of HAT-P-4 and Kronos-Krios occurred in the absence of engulfment because of large uncertainties on abundances that anchor the upper and lower portions of the $T_c$ trend. To test this, we simulated 1000 systems assuming the HAT-P-4 and Kronos-Krios companion masses and convective zones, but with abundances drawn from Gaussian distributions with widths equal to the average abundance scatter per element between the companions of our ten twin systems excluding HAT-P-4 (Figure \ref{fig:figure8}, right panel, blue points). There are 33 simulated systems with $\Delta$ln($Z$) values that exceed that of HAT-P-4 ($\Delta$ln($Z$) $>$ 1.82), of which two also have recovered amounts of engulfed mass greater than HAT-P-4's value of 5.60 $M_{\oplus}$. However, there are no simulated systems with $\Delta$ln($Z$) or recovered amounts of engulfed mass greater than those of Kronos-Krios (Figure \ref{fig:figure10}). We conclude that the HAT-P-4 $T_c$-dependent abundance pattern can occur randomly in the absence of engulfment, but not that of Kronos-Krios. Thus, Kronos-Krios may be a true engulfment detection whereas HAT-P-4 is likely not.

The lack of clear engulfment detections in our sample can be explained by our \texttt{MESA} analysis (Behmard et al. \emph{in review}), which predicts that observable refractory enrichments from 10 $M_{\oplus}$ engulfment events occurring at ZAMS will become depleted on timescales of $\sim$2 Myr$-$2 Gyr for solar-like (0.8$-$1.2 $M_{\odot}$) stars. The largest and longest-lived signatures are exhibited by 1.1$-$1.2 $M_{\odot}$ stars ($\sim$2 Gyr). We thus recommend these stars as the best candidates for engulfment detections. We also considered other engulfment scenarios assuming a 1 $M_{\odot}$ star, and found that engulfment signature timescales increase to $\sim$1.5 Gyr for late-stage (300 Myr$-$3 Gyr post-ZAMS) engulfment, and $\sim$3 Gyr for sub-solar ($Z$ = 0.012) engulfing star metallicities. Most ($\sim$85\% within mass measurement error) of the stars composing the 29 binaries in our sample assessed for engulfment signatures are in the solar-like mass range. In addition, there are only two systems younger than 2 Gyr (HD 202772 and WASP-180), and only 1 system younger than 3 Gyr with sub-solar metallicities (Kepler-477). Thus, the timescales of observable signatures from nominal 10 $M_{\oplus}$ engulfment are short compared to the system lifetimes. Our \texttt{MESA} results also show that refractory enhancements exhibit half-lives of $\sim$6$-$500 Myr (Figure \ref{fig:figure4}). This implies that unless the engulfment event happened recently, we can only recover clear engulfment signatures by taking observations soon after the engulfment event. 
Perhaps this is the case for Kronos-Krios assuming it is a true engulfment detection.

\begin{figure*}[t]
    \centering
        \includegraphics[width=0.7\textwidth]{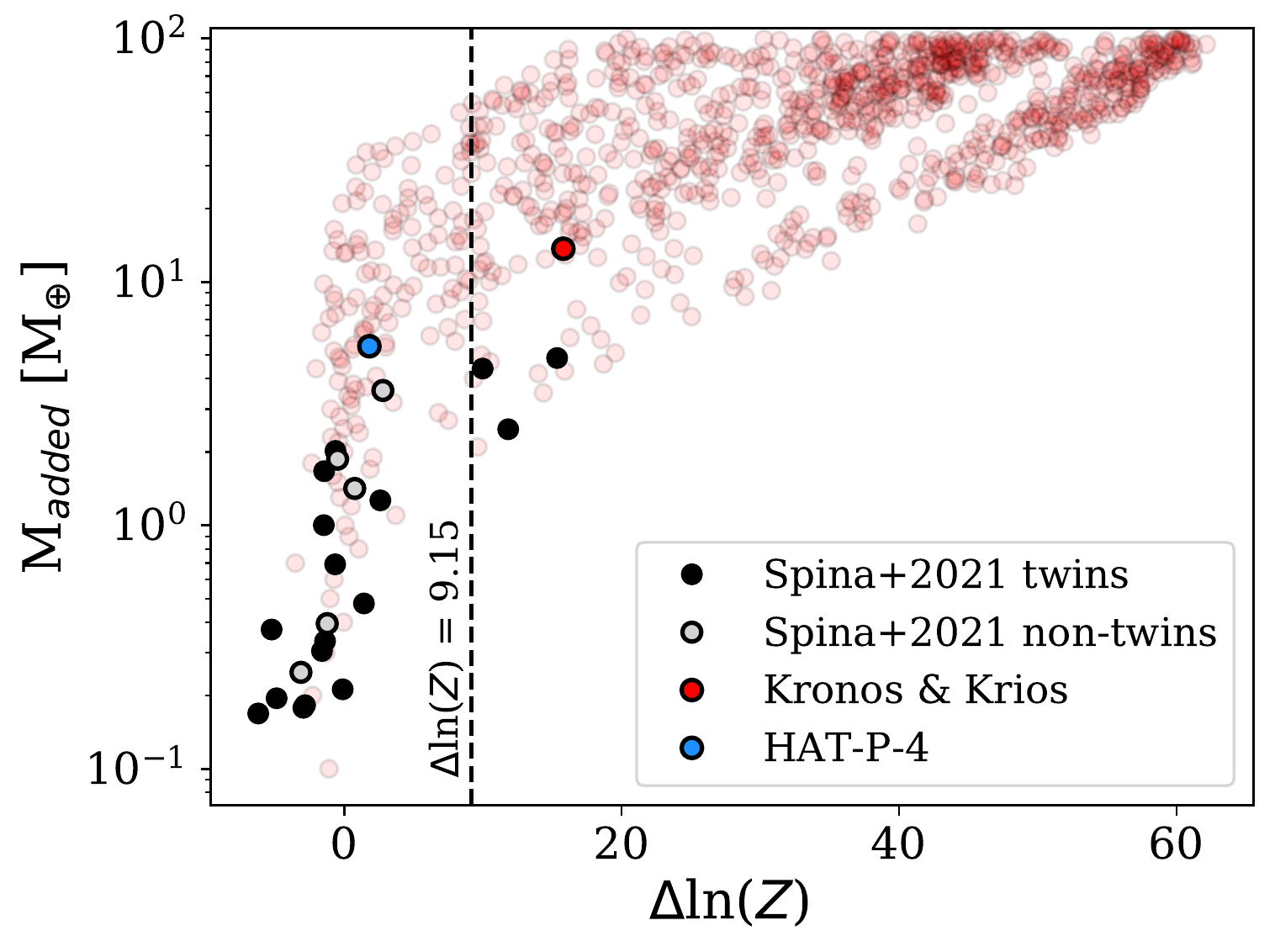} 
    \caption{Estimated amounts of engulfed material from our engulfment model fits vs. $\Delta$ln($Z$) values for the ``chemically anomalous" systems reported \citet{spina2021}. The twin ($\Delta$$T_{\textrm{eff}}$ $<$ 200 K, \citealt{andrews2019}) pairs are shown in black while the non-twin pairs are shown in gray. The points corresponding to 0.1$-$100 $M_{\oplus}$ simulated engulfment systems are represented by the transparent red dots in the background for comparison. Kronos-Krios and HAT-P-4 are also shown for comparison as the red and blue dots, respectively. The maximum $\Delta$ln($Z$) value for the synthetic non-engulfment distribution of 9.15 is marked by the dashed line.} 
\label{fig:figure11}
\end{figure*}

Our \texttt{MESA} results also underscore the importance of using stellar twin binaries for planet engulfment surveys. Refractory depletion rates vary as a function of engulfing star mass and spectral type, even in the absence of planet engulfment (\citealt{sevilla2022}, Figure 9). Thus, non-twin stellar siblings will always exhibit different photospheric abundances. As mentioned in Section \ref{sec:sample}, only eleven of the 36 binaries in our sample qualify as twins. This is another potential contributing factor to our lack of engulfment detections. We thus recommend that future engulfment surveys focus solely on stellar twin systems. Considering the eleven twin systems in our sample, we calculated an upper limit engulfment detection rate for our study using the observable signature timescales from our \texttt{MESA} analysis. This rate was taken as the average in log space of signature timescales (which varies as a function of engulfing star mass) over system age ratios for the eleven twin systems. The resulting rate is $\sim$4.9\%, though we note that the true rate will be much lower since it should be multiplied by a factor corresponding to the intrinsic engulfment rate, which is unknown.

Our results are in contradiction with previous studies that report high rates of engulfment detections. For example, \citet{spina2021} claim an engulfment rate of $\sim$20$-$35\% for their sample of 107 binary systems. They based this on a large fraction of systems (``chemically anomalous" pairs) with high [Fe/C] ratios, $\Delta$[Fe/H], and $\Delta$A(Li). No other abundances were examined and thus there is no analysis of $T_{c}$ trends. In addition, the elemental abundances of the 107 systems were derived from multiple literature sources that took observations with different instruments, and employed different spectral synthesis pipelines and line lists \citep{desidera2004,desidera2006,hawkins2020,nagar2020}. Such heterogeneous methods can introduce systematic bias into abundance samples (e.g., \citealt{schuler2011, liu2018}). Finally, many of the binaries employed in this study do not qualify as stellar twins; \citet{spina2021} imposed a $\Delta T_{\textrm{eff}}$ cutoff of 600 K. Thus, we argue that \citet{spina2021} lack sufficient evidence for their $\sim$20$-$35\% engulfment rate claim.

\citet{spina2021} based their claim on 33 ``chemically anomalous" pairs among their total sample of 107 binaries. Eleven of these 33 pairs were observed by \citet{spina2021} with the HARPS spectrograph and analyzed with \texttt{MOOG}. The abundance measurements of the remaining pairs are drawn from other catalogs; another eleven systems are from \citet{hawkins2020}, four from \citet{desidera2006}, two from \citet{nagar2020}, and one from \citet{desidera2004}. The last four systems are included in our sample (Kronos-Krios, HAT-P-4, 16 Cygni, XO-2). As discussed earlier, we only consider Kronos-Krios and HAT-P-4 as potential engulfment detections. We assessed the other ``chemically anomalous" pairs as follows. The \citet{desidera2004}, \citet{desidera2006}, and \citet{nagar2020} studies do not provide abundances beyond Fe, but \citet{spina2021} and \citet{hawkins2020} measured a set of abundances spanning a wide range of $T_c$ (e.g., C, N, Mn, Cr, Si, Fe, Mg, Ni, V, Ca, Ti, Al, Y). We analyzed the 22 \citet{spina2021} and \citet{hawkins2020} ``chemically anomalous" pairs with our engulfment model considering the abundances listed above. The fitted amounts of engulfed material from our engulfment model vs. $\Delta$ln($Z$) values for these pairs are shown in Figure \ref{fig:figure11}, with twin ($\Delta$$T_{\textrm{eff}}$ $<$ 200 K, \citealt{andrews2019}) pairs represented by black dots and non-twin pairs represented by gray dots. Three systems exhibit $\Delta$ln($Z$) values above 9.15, the maximum $\Delta$ln($Z$) of our synthetic non-engulfment systems. All three systems qualify as binary twins. They have 2.48$-$4.87 $M_{\oplus}$ fitted amounts of engulfed material from our engulfment model, and $\Delta$ln($Z$) values ranging from 10.0$-$15.4. We conclude that these three systems are potential engulfment detections. 

We carried out a similar analysis to estimate the mass of engulfed material for the remaining seven \citet{desidera2004}, \citet{desidera2006}, and \citet{nagar2020} ``chemically anomalous" pairs considering just Fe and its abundance in bulk Earth compositions, and estimated 1.27$-$12.34 $M_{\oplus}$ amounts of engulfed material. 
Only three of these seven systems qualify as twins. If we consider just the twin pairs, the amount of engulfed material drops to 1.27$-$2.50 $M_{\oplus}$, and there are no $\Delta$ln($Z$) values to provide further evidence for these systems as engulfment detections. We thus conclude that there are only five potential detections (Kronos-Krios, HAT-P-4, and three additional systems) in the \citet{spina2021} sample of 107 systems, yielding an engulfment rate of $\sim$4.7\%. However, Kronos-Krios, HAT-P-4, 16 Cygni, and XO-2 were likely included because they are reported as possible engulfment detections in previous studies (Table \ref{tab:table1}). If we remove these systems, there are only three potential detections out of 103 systems, yielding an engulfment rate of $\sim$2.9\%. This is much lower than the $\sim$20$-$35\% \citet{spina2021} engulfment rate claim. 

We conclude that engulfment detections are rare, and put forward the possibility that the abundance differences of HAT-P-4 are primordial. Those of Kronos-Krios may also be primordial, but there is evidence that this system is a true engulfment detection because its strong $T_c$ trend is not produced randomly from large uncertainties on low and high $T_c$ abundances. Considering the HAT-P-4 case, if large ($\Delta$[X/H] $>$ 0.05 dex) primordial abundance differences between binary companions are common, it may not be safe to assume that stellar siblings born from the same molecular cloud are always chemically homogeneous. This could undermine the validity of galactic archaeological tools used to trace stars back to their parent clouds, namely chemical tagging. There are hints that chemical tagging may have limitations. As mentioned in Section \ref{sec:potential_detections}, \citet{ness2018} found a small population of chemically inhomogeneous red giant pairs in open clusters. Similarly, the \citet{hawkins2020} study of 25 wide binaries reported that while 80\% are homogeneous to 0.02 dex levels, six pairs exhibit $\Delta$[Fe/H] $>$ 0.05 dex. Larger wide binary pair samples could be used to place upper limits on abundance differences as a function of separation, and may be aided by ongoing high-resolution spectroscopic surveys such as APOGEE \citep{abdurrouf2022} and GALAH \citep{buder2021}.

\section{Summary} \label{sec:summary}
We carried out a Keck-HIRES survey of 36 planet host binaries and examined their differential stellar abundances for evidence of planet engulfment. However we reiterate that only eleven of these 36 binaries qualify as stellar twins ($\Delta$$T_{\textrm{eff}}$ $<$ 200 K, \citealt{andrews2019}), and our \texttt{MESA} results show that reliable engulfment signatures can only be detected in twin systems because refractory depletion rates vary as a function of engulfing star type. None of the systems in our sample exhibit clear engulfment signatures, which dovetails with our \texttt{MESA} results that show observable signatures in solar-like (0.8$-$1.2 $M_{\odot}$) stars are depleted below observable levels ($\Delta$[X/H] $>$ 0.05 dex) within $\sim$2 Gyr after the engulfment event (Behmard et al. \emph{in review}). Only one of our twin binary systems, HD 202772, has an age below 2 Gyr (1.8 Gyr, \citealt{wang2019}).

Among our twin systems, only HAT-P-4 exhibits a possible engulfment signature. If engulfment occurred in this system, it must have happened within the last 2 Gyr, which is less than half of HAT-P-4's estimated age (4.2 Gyr, \citealt{ment2018}). This makes the engulfment scenario somewhat unlikely. Alternatively, HAT-P-4's abundance differences may be primordial as evidenced by the large projected separation (30,000 $\pm$ 140 AU) between the binary companions. This projected separation is larger than that of any other system in our sample by an order of magnitude (Table \ref{tab:table7}). Similarly, we suggest that the Kronos-Krios abundances differences may be primordial based on the large projected separation of the system (11,000 $\pm$ 12 AU).

We used our engulfment model to analyze previously published datasets for ten planet host binary systems (HAT-P-1, HD 20781-82, XO-2, WASP-94, HAT-P-4, HD 80606-07, 16-Cygni, HD 133131, HD 106515, WASP-160; Table \ref{tab:table1}), of which four to six are claimed as engulfment detections depending on the study. None of the systems can be claimed as detections according to our criteria for engulfment, outlined in Section \ref{sec:potential_detections}. We also examined how abundance scatter depends on line lists employed in spectral synthesis pipelines, and found that abundance precision increases with larger numbers of strong spectral features per chemical species (Figure \ref{fig:figure8}, left panel). Elements with low $T_c$ (e.g., volatiles such as C, N, O), and high $T_c$ (e.g., Y) lack an abundance of strong features and thus exhibit large scatter. Because these abundances are important for anchoring $T_c$ trends, we conclude that $T_c$ patterns can randomly result from poorly measured abundances in the absence of engulfment (Figure \ref{fig:figure8}, right panel). We tested if the HAT-P-4 and Kronos-Krios $T_c$ trends can be randomly produced from large uncertainties on low and high $T_c$ abundances, and found that this is the case for HAT-P-4, but not Kronos-Krios. We conclude that Kronos-Krios may still be a true engulfment detection, but HAT-P-4 is likely not.

Our results contradict previous studies that report high rates of engulfment, namely \citet{spina2021} that claimed an engulfment rate of $\sim$20$-$35\% for their sample of 107 binary systems. We analyzed the abundance patterns of their ``chemically anomalous" systems with our engulfment model, and determined that the true engulfment rate is closer to $\sim$2.9\%. This is comparable to the upper limit engulfment detection rate we calculated from our $\texttt{MESA}$ engulfment signature timescales ($\sim$4.9\%). Our results suggest that reported detections of planet engulfment may instead be due to primordial chemical differences between stellar companions. To confirm this, the homogeneity of bound stellar siblings as a function of binary separation should be investigated further in future studies.

\begin{acknowledgments}
We thank Jim Fuller and Heather Knutson for insightful comments that improved the final manuscript. A.B. acknowledges funding from the National Science Foundation Graduate Research Fellowship under Grant No. DGE1745301. This work benefited from involvement in ExoExplorers, which is sponsored by the Exoplanets Program Analysis Group (ExoPAG) and NASA’s Exoplanet Exploration Program Office (ExEP).
\software{\texttt{numpy} \citep{numpy}, \texttt{matplotlib} \citep{matplotlib}, \texttt{pandas} \citep{pandas}, \texttt{scipy} \citep{scipy}, \texttt{astropy} \citep{astropy:2013, astropy:2018}, \texttt{dynesty} \citep{speagle2020}}
\end{acknowledgments}

\appendix
\setcounter{figure}{0}                       
\renewcommand\thefigure{A.\arabic{figure}}

\section{Abundance Error Analysis} \label{sec:snr_test}

We obtained Keck-HIRES observations of eight bright stars (HIP 38931, HIP 44137, HIP 47288, HIP 16107, HIP 14300, HIP 15099, HIP 14241, HIP 21272) at five different SNR levels, and calculated the variance in their \texttt{SME} abundance predictions. These eight stars span the $T_{\textrm{eff}}$ range of our engulfment sample, and a wide [Fe/H] range of $-$0.39 to +0.37 dex, making the results of this test relevant for a diverse set of stars. We collected 3--6 spectra per star and SNR level, and found that the variance in measured abundances is $\lesssim$0.03 dex for refractory species, and higher for volatile species with variance up to $\sim$0.1 dex. As expected, the variance decreases dramatically as a function of SNR for all abundances and stars, with the exception of HIP 38931 which exhibits large scatter in the volatile abundances even as the highest SNR level (200/pix) is approached. This is likely due to its low temperature ($T_{\textrm{eff}}$ = 4680 K) and low metallicity ([Fe/H] = $-$0.17 dex) which together create a favorable environment for forming volatile-bearing molecular species whose spectral features are difficult to model with \texttt{SME}. Because of this, we consider \texttt{SME}-determined abundances for targets with $T_{\textrm{eff}}$ $<$ 4700 K and sub-solar metallicities to be suspect. 

\begin{figure}[t!] 
\centering 
    \vspace*{0.3in}
    \includegraphics[width=0.475\textwidth]{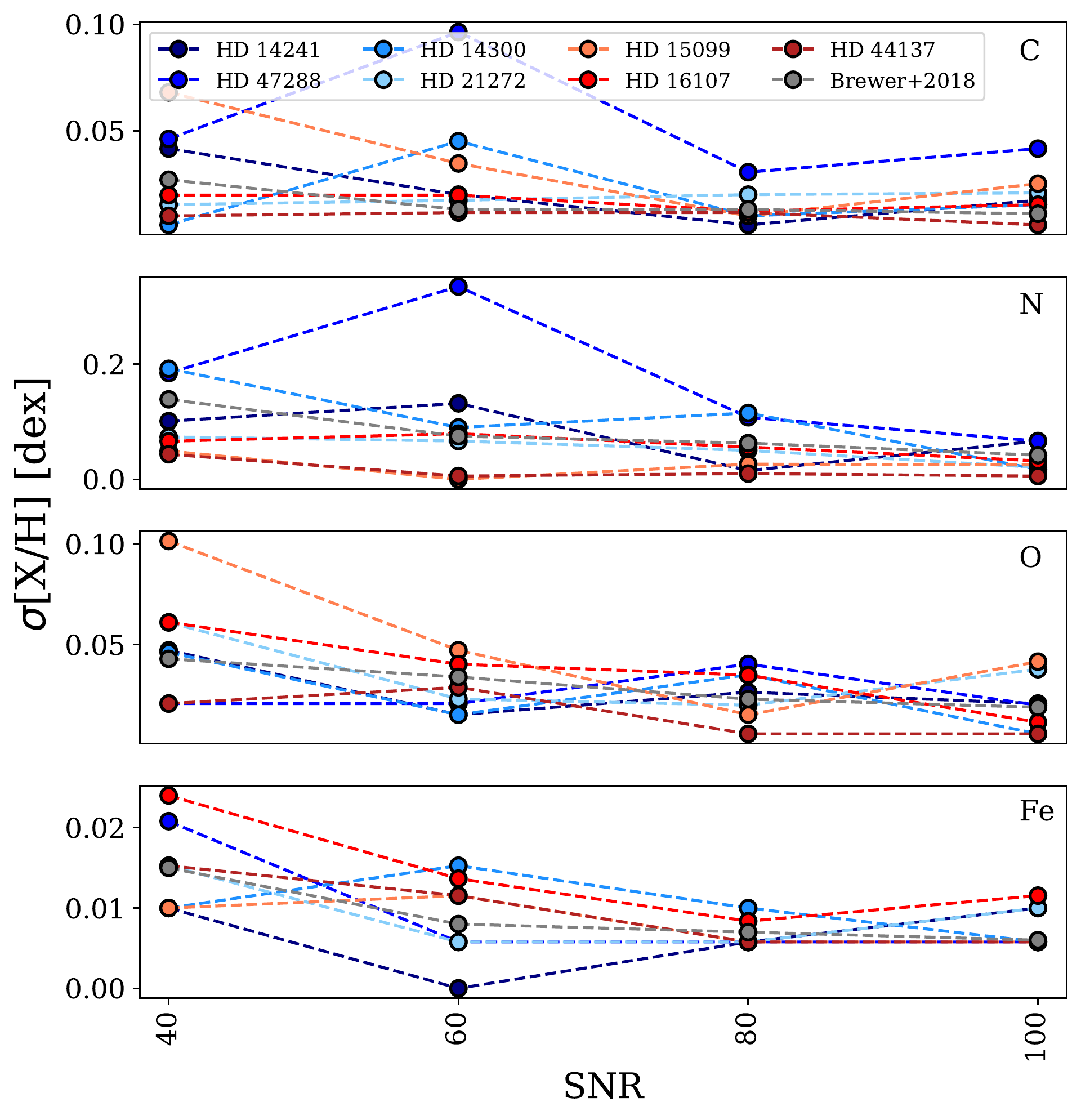}
\caption{Standard deviation in \texttt{SME} abundance predictions from multiple HIRES observations of seven bright stars (colored circles), and the \citet{brewer2018} scatter in \texttt{SME} abundance predictions for a solar spectrum with varying amount of added Gaussian random noise to mimic varying SNR levels (gray). The seven bright stars are colored in order of increasing metallicity (dark blue to dark red). The abundances displayed are C, N, O, and Fe from top to bottom, as a function of SNR = 40, 60, 80, and 100.}
\label{fig:figureA}
\end{figure}

Excluding HIP 38931, we found that the abundance scatter of the remaining seven bright stars agrees with the abundance errors reported in \citet{brewer2018}. To illustrate this, we plotted the standard deviation of C, N, O, and Fe \texttt{SME} abundance predictions for these seven stars against the \citet{brewer2018} solar spectra abundance scatter for SNR levels of 40, 60, 80, and 100 in Figure \ref{fig:figureA}. We chose these abundances because spectral synthesis codes like \texttt{SME} struggle to model the features of volatile species like C, N, and O due to molecular lines, and Fe provides a good comparison point by possessing many easily modeled lines. As expected, the abundance scatter trends as a function of SNR are approximately monotonic, though the scatter of HIP 47288, HIP 14300, and HIP 14241 noticeably deviate for C and N. This is likely because these are the most metal poor stars remaining in our now seven bright star sample, and are thus more likely to host volatile-bearing molecules in their photospheres. 

It can be seen that the \citet{brewer2018} predictions (gray circles) well-represent the abundance scatter across all seven stars that we observed (colored circles). The average absolute difference between the abundance scatter reported by \citet{brewer2018} and those of our seven bright stars is $\sim$0.012 dex, and the \citet{brewer2018} scatter is larger $\sim$57\% of the time across all SNR levels and abundances. Thus, the \citet{brewer2018} abundance scatter is a good approximation of \texttt{SME} abundance errors for our engulfment sample, and we derived our errors by linearly interpolating through SNR in Table 2 of \citet{brewer2018} to match the individual SNR of each star (Table \ref{tab:table5}).

\bibliography{mybib}{}
\bibliographystyle{aasjournal}



\end{document}